\documentclass[12pt]{article}

\usepackage{lscape,multirow}
\usepackage{graphics, amsmath, amssymb, multirow, mathrsfs, graphicx, bm, colordvi, verbatim}
\usepackage{setspace, natbib}
\usepackage{caption, tabularx}
\usepackage{enumitem,colortbl}
\usepackage{url}
\usepackage{tablefootnote, ragged2e}
\usepackage{mathptmx}
\usepackage{pdflscape}
\usepackage{amsthm}
\usepackage{chngcntr}
\usepackage{changepage}
\usepackage{tikz}
\usepackage{tgpagella}
\usepackage{longtable}
\usepackage{pdflscape}
\usepackage{afterpage}
\usepackage{hyperref}
\usepackage{graphicx}
\usepackage[stable]{footmisc}
\usepackage{array} % if not already included
\newcolumntype{P}[1]{>{\centering\arraybackslash}p{#1}}
\hypersetup{
    colorlinks=true,       % false: boxed links; true: colored links
    linkcolor=blue,          
    citecolor=blue,        
    filecolor=magenta,      
    urlcolor=blue           
}

\usepackage{booktabs}

\usepackage{dcolumn}
\newcolumntype{d}[1]{D..{#1}} % for alignment of numbers on decimal marker

\usepackage[normalem]{ulem}

\newcolumntype{Y}{>{\centering\arraybackslash}X}

\usepackage{siunitx}
\usepackage[left=1in,right=1in,top=1in, bottom=1in]{geometry}
\usepackage{setspace}
\renewcommand{\thefootnote}{\fnsymbol{footnote}}

\begin{document}
\newgeometry{left=1in,right=1in,top=0in, bottom=1in} % adjust the value as needed
\begin{titlepage}
\title{\bf Chronologically Consistent Large Language Models\thanks{\scriptsize Songrun He is at Washington University in St. Louis (\url{h.songrun@wustl.edu}). Linying Lv is at Washington University in St. Louis (\url{llyu@wustl.edu}). Asaf Manela is at Washington University in St. Louis (\url{amanela@wustl.edu}). Jimmy Wu is at Washington University in St. Louis (\url{jimmywu@wustl.edu}).}}
\author{Songrun He \quad\quad Linying Lv \quad\quad Asaf Manela \quad\quad Jimmy Wu}
        \date{First draft: February 2025. This draft: July 2025.}
        \maketitle
%         1. We estimate the lookahead bias in an asset pricing application.
%         2. Training a chronologically consistent LLM requires a large enough data set that allows the language model to be accurate despite using less data / training a chronologically consistent LLM is trivial, but training one that competes with state-of-the-art LLMs is challenging.
%         3. Lookahead bias is application-specific -- even though our LLM does worse than llama3 on sentence completion tasks, it performs similarly on an asset pricing task because the predictive model layered on top of the LLM can adjust.
%         4. Overall, we provide a constructive solution to address the concern of lookahead bias in economics and finance, and more broadly in the realm of social sciences. 

{\begin{abstract}\onehalfspacing
   Large language models are increasingly used in social sciences, but their training data can introduce lookahead bias and training leakage. A good chronologically consistent language model requires efficient use of training data to maintain accuracy despite time-restricted data. Here, we overcome this challenge by training a suite of chronologically consistent large language models, ChronoBERT and ChronoGPT, which incorporate only the text data that would have been available at each point in time. Despite this strict temporal constraint, our models achieve strong performance on natural language processing benchmarks, outperforming or matching widely used models (e.g., BERT), and remain competitive with larger open-weight models. Lookahead bias is model and application-specific because even if a chronologically consistent language model has poorer language comprehension, a regression or prediction model applied on top of the language model can compensate. In an asset pricing application predicting next-day stock returns from financial news, we find that ChronoBERT and ChronoGPT's real-time outputs achieve Sharpe ratios comparable to a much larger Llama model, indicating that lookahead bias is modest. Our results demonstrate a scalable, practical framework to mitigate training leakage, ensuring more credible backtests and predictions across finance and other social science domains.%\asaf{can we distinguish lookahead from survivorship / training leakage?}
\end{abstract}}\small~\\
        {\bf JEL Classification}:  G11, G12, G17\\
        {\bf Keywords}: Large language model, chronological consistency, lookahead bias, training leakage, backtesting

        \thispagestyle{empty}
\end{titlepage}
    \restoregeometry

    \newpage
    \setcounter{page}{1}
    \renewcommand{\thefootnote}{\arabic{footnote}}

    \setstretch{1.5}
    \doublespacing

% \newpage
% \songrun{
% \begin{enumerate}
% \item Sentence transformer experiments: (1) ModernBERT on the original dataset; (2) ModernBERT on the news headline dataset; (3) ChronoBERT on the news headline dataset.
% \item Consider adding (1) Sarkar and Vafa (2024) experiment of predicting risk/election outcome and (2) Glasserman and Lin (2023) experiment of masked language modeling in a later draft.
% \end{enumerate}
% }

\newpage
\begin{quote}
    ``Obviously, the time continuum has been disrupted, creating a new temporal event sequence resulting in this alternate reality.''
    
    \hfill -- Dr. Brown, \textit{Back to the Future Part II} % https://www.youtube.com/watch?v=GfmdW3hiu8w
\end{quote}

\section{Introduction}\label{sec:intro}

Large language models (LLMs) now permeate economics and finance, revealing nuanced patterns in unstructured text \citep{HobergManela2025}. They allow researchers to test hypotheses once considered unquantifiable. Yet these models often learn from data that did not exist at the historical moment. This ``lookahead bias'' (or training leakage) undermines empirical findings in settings that require real-time information \citep{glasserman_assessing_2023, sarkar_lookahead_2024, levy_caution_2024, ludwig2025large}.

We address this challenge by training chronologically consistent LLMs trained exclusively on historical textual data available at the time. While training language models with historical data timestamped at the point of its availability is conceptually straightforward, ensuring these models are competitive with highly capable open-weight counterparts remains a significant challenge.

Building such models requires overcoming two primary obstacles: they require large amounts of computational power, and they rely on limited historical text. To tackle the computational side, we draw on efficient training methods \citep{warner2024smarter, modded_nanogpt_2024} to lower computing costs. To maximize information gained from a limited corpus, we follow \citet{gunasekar2023textbooks} by selecting diverse, high-quality data, carefully filtered by publication date. This two-pronged strategy yields strong, chronologically consistent models even under tight resource constraints.

Our chief contribution is \emph{ChronoBERT} and \emph{ChronoGPT}: a pair of models for each year starting in 1999 that never see future data during training. On the GLUE benchmark \citep{wang2018glue}, Both ChronoBERT and ChronoGPT outperform existing no-leakage models, including StoriesLM \citep{sarkar_storieslm_2024} and FinBERT \citep{huang2023finbert}. Even our earliest ChronoBERT matches or surpasses the popular BERT \citep{devlin-etal-2019-bert}, which ranks first in downloads among all language models on Hugging Face as of February 2025.\footnote{Based on download statistics from Hugging Face at \url{https://huggingface.co/models?sort=downloads}.} Strong language understanding, we show, does not require data from the future. While ChronoBERT and ChronoGPT emerge as the strongest leak-free LLMs to date, they also maintain a modest scale, making them feasible for large embedding tasks where massive models (e.g., Llama) are too costly.

Among all social sciences, finance is particularly sensitive to lookahead bias.
Market efficiency tests \citep{fama1970efficient} assume that prices reflect only known facts at the time. A model that ``time-travels'' by reading tomorrow’s news distorts such tests. While researchers could keep a held-out sample of recent data to avoid lookahead bias, the limited panel of asset prices commonly studied forces most studies to rely on backtesting. Hence a chronologically inconsistent language model can bias measures of risk and market inefficiency.

% Beyond chronological consistency, predicting stock returns in fairly efficient capital markets requires two additional model qualities: (i) good language understanding to interpret the news, and (ii) up-to-date knowledge to frame news in their broader context as it is understood at the time.\asaf{we may add something like this to explain why we need the series of models more recent that chrono-bert-1999} Therefore, just like tests of market efficiency are a joint-test that the market is efficient and the pricing model is correct, tests for lookahead bias require a language model with both qualities.

We test for lookahead bias by forecasting stock returns from news with ChronoBERT and ChronoGPT. Leveraging extensive financial newswire data, we find that the portfolio performance of ChronoBERT and ChronoGPT matches that of Llama 3.1 \citep{dubey2024llama}, with both models delivering economically substantial and statistically significant gains compared to StoriesLM \citep{sarkar_storieslm_2024} and FinBERT \citep{huang2023finbert}. Our findings suggest that lookahead bias in this context is relatively modest.

An important observation is that the impact of lookahead bias is model- and application-specific. While ChronoBERT and ChronoGPT may exhibit lower language comprehension on general tasks compared to unconstrained models, downstream predictive models built on top of our models can adapt to these limitations, mitigating potential drawbacks in financial forecasting.

Our contribution is threefold. First, we explain how to build chronologically consistent LLMs that preserve validity in economic and financial tests. Second, we find that strong language understanding can be attained without introducing training leakage: ChronoBERT and ChronoGPT remain competitive with powerful, open-weight baselines. Third, we measure lookahead bias in news-based return forecasting and find it modest. These points open a path to more credible LLM applications across the social sciences.

Our work is related to two broad strands of literature. First, a large literature studies how news forecasts stock returns \citep[e.g.,][]{tetlock_more_2008, jiang2021pervasive, ke_predicting_2019}. Early dictionary-based or word-count methods measured only coarse signals, delivering smaller economic effects than modern LLMs.
LLMs have pushed text-based stock return forecasting further. \citet{chen2023expected} show that news embeddings from LLMs strongly predict next-day returns. \citet{lopez-lira_can_2023} demonstrate how prompting LLMs yields robust signals. These results highlight the growing potency of advanced NLP in finance.

Our main contribution to this literature is to show that the robust return predictability achieved through LLMs is not driven by lookahead bias, thereby addressing a critical concern in the use of these advanced models for financial forecasting.

A second strand of literature pertains to the application and development of natural language processing (NLP) tools for financial economics research \citep{HobergManela2025}. Methodological advancements in this area have evolved from early dictionary-based approaches \citep{tetlock_giving_2007, loughran2011liability}, to text regressions \citep{manela2017news, kelly2021text}, to topic modeling \citep{bybee2024business}, to neural embedding methods \citep{goetzmann2022crash}, and most recently, to the integration of LLMs \citep{ Jha_Liu_Manela_2020_goodfin, chen2023expected, lv2024value}. These developments underscore the growing demand for more advanced and scalable tools to answer research questions in finance and economics.

In this broader context, our main contribution is a framework for developing LLMs that are both free of lookahead bias and capable of high-level language comprehension.  Our approach does not require masking \citep{glasserman_assessing_2023, engelberg_entity_2025}, which can destroy information, and may not completely remove the look-ahead bias of high-memory LLMs \citep{levy_caution_2024, lopez-lira_memorization_2025, didisheim_ais_2025}. Instead, we train a series of chronologically consistent models with knowledge cutoffs from 1999 to 2024 that offer superior language understanding and more up-to-date knowledge than similar attempts \citep[e.g.,][]{sarkar_storieslm_2024}.\footnote{In contemporaneous work, \citet{rahimikia2025r} introduce a series of LLMs trained on time-stamped proprietary financial texts for predictions. They do not evaluate the models' general language understanding capabilities, but based on the financial forecasting metrics they do provide, their best-performing model generates a long-short portfolio Sharpe ratio of 3.45 from 2017 to 2023, significantly underperforming ChronoGPT and ChronoBERT. Unfortunately, the authors have since retracted their models from Hugging Face, preventing further direct comparison.}

The rest of the paper is organized as follows: Section \ref{sec:method} describes our approach and data for model pretraining and evaluation. Section \ref{sec:empirical} presents the empirical performance of our model. Section \ref{sec:conclusion} concludes.

\section{Methodology and Data}\label{sec:method}

In this section, we outline the methodology we use for model pretraining and describe the approach for evaluating their performance. Specifically, we assess their ability in both language understanding and asset pricing tasks.

\subsection{Pretraining Methodology for ChronoBERT}

When pretraining ChronoBERT, we incorporate a modern BERT architecture from \citet{portes2023mosaicbert} and \citet{warner2024smarter}.\footnote{We thank the authors for providing their pretraining code at \url{https://github.com/mosaicml/examples/tree/main/examples/benchmarks/bert} and \url{https://github.com/AnswerDotAI/ModernBERT}.} Compared to the original BERT model by \citet{devlin-etal-2019-bert}, this enhanced architecture integrates recent advancements in rotary positional embeddings that support longer context lengths and employ flash attention, significantly improving pretraining efficiency and computational speed.

For the pretraining task, we follow \citet{warner2024smarter} by adopting masked token prediction while omitting the next sequence prediction task, as prior research has shown the latter increases training overhead without meaningful performance gains.\footnote{We provide details on ChronoBERT's masked language modeling pretraining objective and attention mechanisms in Appendix \ref{sec:architectures} and contrast them with those of ChronoGPT.}

The quality of pretraining data is critical to achieving BERT-level performance. \citet{gunasekar2023textbooks} demonstrate that using high-quality, ``textbook-like" data leads to faster convergence and improved model outcomes. Motivated by this insight, we filter our pretraining corpus using the FineWeb-edu classifier from \citet{penedo2024fineweb}, retaining only texts with scores above two.\footnote{We thank the authors for providing the classifier at \url{https://huggingface.co/HuggingFaceFW/fineweb-edu-classifier}.}

However, restricting the corpus to texts with historical dates—particularly from early historical periods—introduces data scarcity challenges. \citet{muennighoff2023scaling} explore the scaling laws of LLMs under data constraints, highlighting the benefits of iterative training on limited high-quality data. Following their insights, we train our model over multiple epochs to maximize learning from the available corpus. Our first model checkpoint $\text{ChronoBERT}_{1999}$ is trained on 460 billion tokens, with more than 70 epochs through the dataset.

\subsection{Pretraining Methodology for ChronoGPT}
ChronoGPT is pretrained using a modified nanoGPT architecture, incorporating key enhancements from \citet{modded_nanogpt_2024}.\footnote{We thank the authors for providing their training framework at \url{https://github.com/KellerJordan/modded-nanogpt}.} Compared to the original GPT-2 implementation by \citet{radford2019language}, this architecture integrates several optimizations, including Rotary embeddings, QK-Norm, skip connection and FlexAttention, improving both computational efficiency and scalability.

ChronoGPT is trained using a causal language modeling (CLM) objective, learning to predict the next token based on previously observed sequences.\footnote{We provide details on ChronoGPT's causal language modeling pretraining objective and attention mechanisms in Appendix \ref{sec:architectures} and contrast them with those of ChronoBERT.} 

To determine the optimal model size, we first conducted a scaling law experiment to assess the impact of parameter count within our data-constrained, chronological training regime. As detailed in Appendix \ref{scaling_law_test}, this test revealed that increasing model size from 124M to 1.5B consistently improved performance on both validation loss and language understanding. Based on these findings, we selected the 1.5B parameter model, as it offered the best performance while remaining within our computational budget.

Given the efficiency gains provided by the modded-nanoGPT optimizer, ChronoGPT requires fewer training iterations to reach competitive performance. We adopt a low-epoch, high-throughput strategy, allowing the model to rapidly extract essential linguistic and contextual structures. Our first vintage model, $\text{ChronoGPT}_{1999}$, is trained on 71 billion tokens over approximately 10 epochs, striking a balance between computational efficiency and representational robustness.

\subsection{Evaluation Methodology}

To evaluate the performance of our models, we assess both language understanding capabilities and economic forecasting performance. We also apply the same evaluation methodology to several other LLMs for benchmarking.

\subsubsection{Language Understanding}

To assess our models' language understanding abilities, we employ distinct evaluations that align with the fundamental architectures of ChronoBERT and ChronoGPT. ChronoBERT is built on an encoder-based structure that creates a bidirectional representation of all input tokens, making it an ideal candidate for classification benchmarks such as GLUE. In contrast, ChronoGPT leverages a generative, autoregressive framework that is naturally suited to open-ended and commonsense reasoning tasks. Accordingly, we compare ChronoBERT against BERT on GLUE and compare ChronoGPT against GPT-2 on HellaSwag, ensuring that each model is assessed on the tasks best suited to its core strengths.

The GLUE evaluation framework, introduced by \citet{wang2018glue}, comprises multiple classification tasks designed to measure a model’s language understanding.\footnote{The details and leaderboard of GLUE evaluation can be found at \url{https://gluebenchmark.com/}. Appendix \ref{sec:glue} provides an overview of the eight GLUE tasks. Appendix \ref{sec:architectures} describes the architectural differences between ChronoBERT and ChronoGPT for sequence classification in GLUE evaluation.} This framework was also the primary evaluation metric used in \citet{devlin-etal-2019-bert} to assess BERT’s language capabilities.

Following pretraining, we further fine-tune the model on task-specific training datasets and evaluate its performance on a held-out test set.

For fine-tuning, we adopt the training specifications and hyperparameters outlined in \citet{warner2024smarter}. Among the eight GLUE tasks, RTE, MRPC, and STS-B are initialized from the MNLI checkpoint to enhance performance.

Beyond GLUE, we evaluate ChronoGPT’s commonsense reasoning with HellaSwag \citep{zellers2019hellaswag} in a zero-shot setting. HellaSwag presents a sentence completion task where each context ends with an incomplete phrase (e.g., ``A woman sits at a piano,'') and the model must choose the most plausible continuation from four possible endings (e.g., ``She sets her fingers on the keys.''). We compute the likelihood of each candidate ending under ChronoGPT’s autoregressive predictions and select the most likely continuation. Since HellaSwag employs adversarial filtering to create highly deceptive distractors, strong performance on this task signals robust contextual and procedural reasoning.\footnote{The leaderboard and data of HellaSwag can be found at \url{https://paperswithcode.com/sota/sentence-completion-on-hellaswag}. Appendix \ref{sec:glue} provides detailed information on the methodology used for HellaSwag evaluations.}

\subsubsection{Predicting Stock Returns using Financial News}

We investigate whether improved language understanding translates into economic gains by using different language models to predict stock returns from economic news. Based on these predictions, we construct portfolios and evaluate the performance of long-short strategies.

Following \citet{chen2023expected}, we first aggregate all news articles' headlines for a stock on a given trading day together. Next, we transform this text into embeddings. Specifically, we process each piece of text through different language models and extract the hidden states of all tokens.\footnote{Different models generate the hidden states in distinct ways, which can impact predictive performance. To account for this, we extract three versions of token embeddings: (1) using the hidden states from the last layer, (2) averaging the hidden states across all layers, and (3) using the hidden states from the first layer. We determine in real time which approach yields the highest `H-L' portfolio Sharpe Ratio over an expanding historical window and use that for forecasting.} The final embedding for each text is obtained by averaging the token embeddings.\footnote{\citet{coleman2020why} provides a rationale for averaging token embeddings to get sequence embeddings.}

% transform each news article's headline into embeddings. Specifically, we process each piece of news through different language models and extract the hidden states of all tokens across all layers. The final embedding for each headline is obtained by averaging the token embeddings.

Next, we link each news article to stock returns on the following trading day and fit a Fama-MacBeth regression with a ridge penalty to map news embeddings to return predictions. Each month $m$, we estimate the following cross-sectional ridge regression:
\begin{equation}
    r_{i, t+1} = \alpha_m + \beta_m' e_{i,t} + \epsilon_{i, t+1}, \quad \text{for }i = 1,\cdots N\text{ and } t=1\cdots T,\footnote{We use a leave-one-out cross-validation algorithm to determine the ridge penalty $\lambda$ chosen from grid points ranging from $10^{-10}$ to $10^{10}$.}
\end{equation}
where $e_{i,t}$ represents of the embedding of all news for firm $i$ on day $t$. To construct real-time out-of-sample forecasts, in month $m'$, we use an average of forecasts over all previous months' cross-sectional models:
\begin{equation}
    \hat{r}_{i,t+1} = \Bar{\alpha}_{m'} + \Bar{\beta}'_{m'}e_{i,t}, \quad \text{for }i = 1,\cdots N\text{ and } t=1\cdots T,
\end{equation}
where $\Bar{\alpha}_{m'} = \frac{1}{m'-1}\sum_{m=1}^{m'-1} \hat{\alpha}_m$ and $\Bar{\beta}_{m'} = \frac{1}{m'-1}\sum_{m=1}^{m'-1} \hat{\beta}_m$.

Using these out-of-sample predictions, we sort stocks into decile portfolios at the end of each trading day, based on forecasts from different language models. We then evaluate the performance of daily-rebalanced long-short decile portfolios constructed from these predictions.

\subsection{Data}

In this section, we present the data we used for pretraining our language models and the financial newswire data we used to evaluate our language models.

\subsubsection{Pretraining Data}

$\text{ChronoBERT}_{1999}$ and $\text{ChronoGPT}_{1999}$ are pretrained for multiple epochs on a corpus of 7 billion tokens comprising chronologically organized English text sourced from diverse domains, including historical web content, archived news articles, and scientific publications. The dataset is fully open-source and is carefully curated to include only high-quality text published before the year 2000, ensuring a focus on no leakage of future knowledge. The final composition of our pretraining corpus was determined through extensive ablation studies to optimize model performance. 

We further conduct incremental training from 2000 to 2024 on a corpus of 65 billion tokens with similar high-quality, diverse, timestamped, and open-source textual data to update knowledge for the model. 

\subsubsection{Financial Newswire Data}

We utilize the Dow Jones Newswire dataset, a real-time newswire subscribed to by major institutional investors. This dataset provides extensive coverage of financial markets, economies, and companies, aggregating reports from leading media sources such as Wall Street Journal, Barron's, and MarketWatch.

The dataset includes news headlines, full article texts, and precise display timestamps, with microsecond-level accuracy for when the news becomes available to clients. Following \citet{ke_predicting_2019}, we focus on firm-specific news that can be attributed to a single company. For each firm-day observation, we aggregate all news headlines related to the firm within the trading day window---spanning from 4:00 p.m. EST on day $t-1$ to 4:00 p.m. EST on day $t$---and treat the combined text as the firm’s textual information. Each concatenated set of headlines is then processed through our embedding framework to generate numerical text representations.

After the embedding step, we merge the news dataset with close-to-close returns on trading day $t+1$ from CRSP together to examine the predictability of stock returns using LLMs trained with real-time available textual data.

Our dataset covers the period from January 2007 to July 2023. The first year serves as a burn-in period to estimate the initial return prediction model, resulting in a final asset pricing test sample spanning January 2008 to July 2023.

\section{Results}\label{sec:empirical}

In this section, we first describe the pretraining process, focusing on validation loss and language understanding as the model is trained on an increasing number of tokens. We then benchmark the language understanding capabilities of our ChronoBERT and ChronoGPT against four other language models. We then validate the chronological consistency of our models. Finally, we assess its asset pricing performance in the news return prediction exercise.

\subsection{Pretraining Model Fit}

\begin{figure}[!htb]
\begin{footnotesize}
\begin{center}
(a) ChronoBERT
\includegraphics[width=\linewidth]{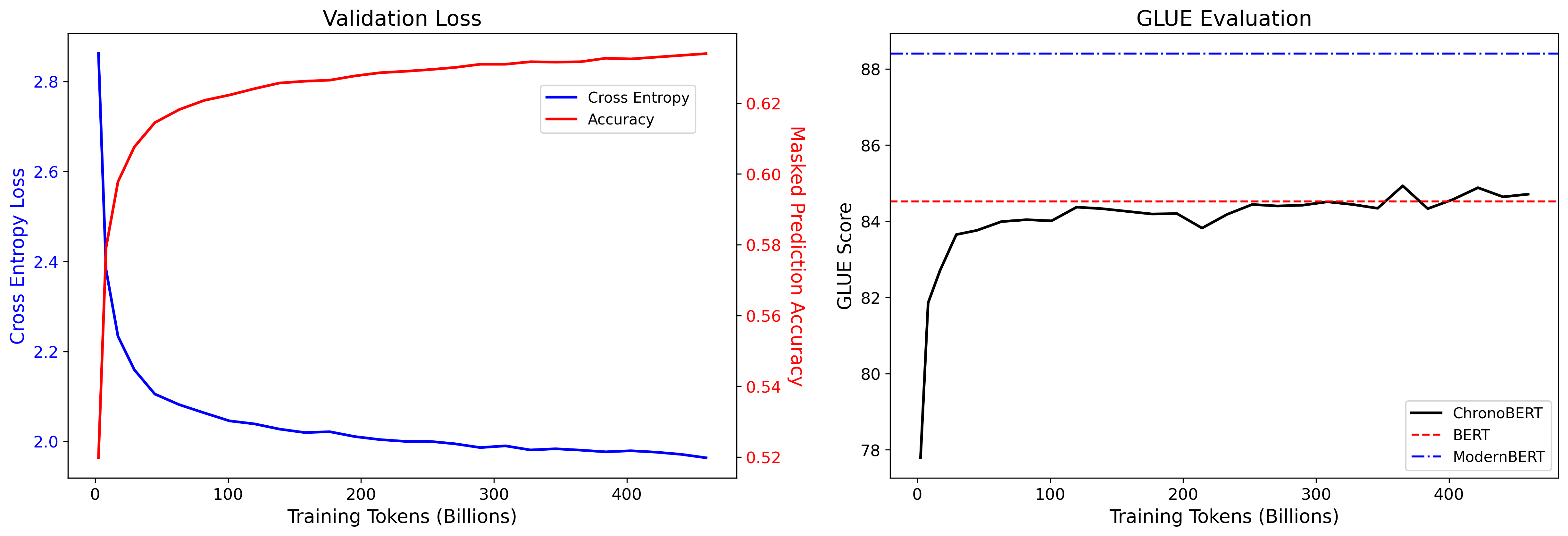}
(b) ChronoGPT
\includegraphics[width=\linewidth]{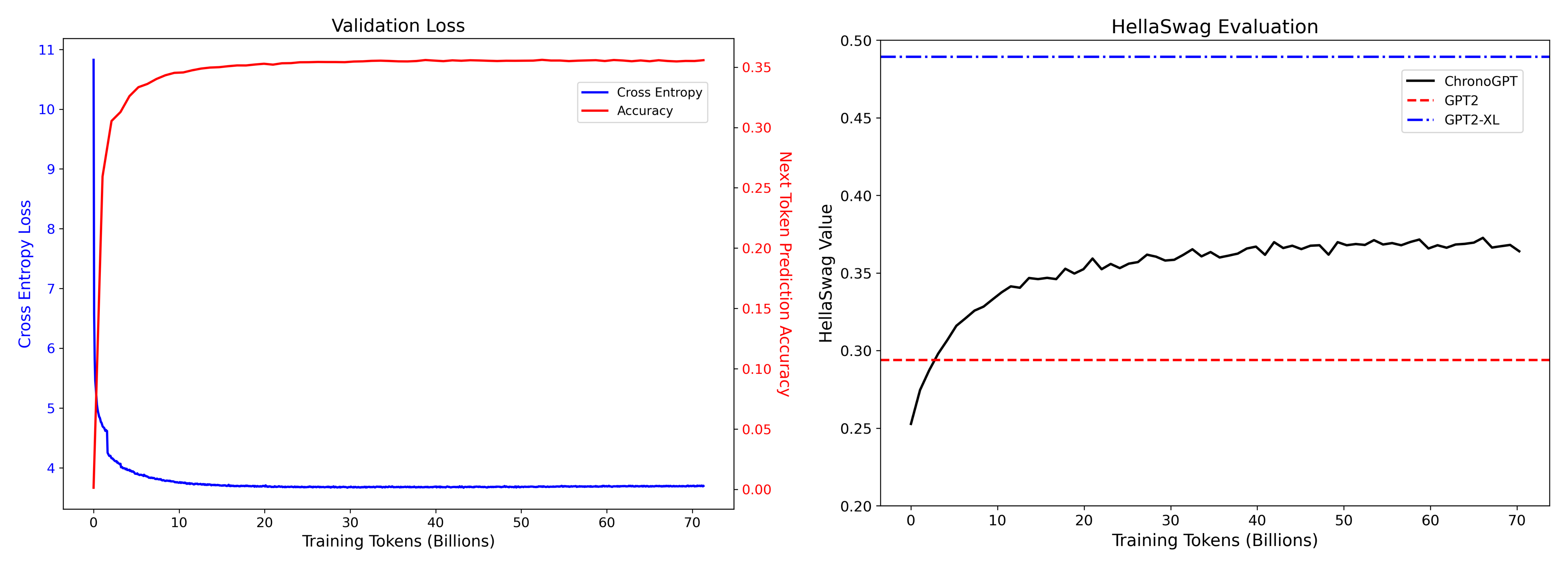}
\end{center}
\end{footnotesize}
\caption{Validation Loss and Evaluation Scores versus Pretraining Tokens}\label{fig:pretrain_chronobert1999}
\bigskip
\small
The left panel shows the validation loss, measured using cross-entropy loss and language prediction accuracy, as the $\text{ChronoBERT}_{1999}$ and $\text{ChronoGPT}_{1999}$ are trained on an increasing number of tokens. The right panel displays the GLUE scores and HellaSwag values as training progresses. For ChronoBERT and ChronoGPT, the final model checkpoint is trained on 460 billion tokens and 70 billion tokens, respectively. The training corpus consists of text up to December 1999.
\end{figure}

Firstly on pretraining performance, our results confirm the scaling law proposed by \citet{muennighoff2023scaling} in a data-limited environment. As shown in Figure \ref{fig:pretrain_chronobert1999}, validation loss (measured via cross-entropy) decreases consistently as the number of training tokens increases.\footnote{For ChronoBERT, we use a subset of the C4 corpus from \url{https://huggingface.co/datasets/allenai/c4} as the validation set. The C4 data is a cleaned version of the Common Crawl data. For ChronoGPT, we adopt the FineWeb validation set, following the setup of \url{https://github.com/KellerJordan/modded-nanogpt}.} Simultaneously, language prediction accuracy improves with training progress.

These improvements translate into enhanced language understanding, as reflected in the GLUE and HellaSwag scores. ChronoBERT surpasses the original BERT's performance on the GLUE benchmark after being trained on approximately 350 billion tokens. Similarly, ChronoGPT outperforms GPT-2 on the HellaSwag benchmark after around 3 billion training tokens.

When evaluated against more demanding benchmarks, the performance of our models begins to plateau. We compare ChronoBERT to ModernBERT, which is trained on a significantly larger text corpus, and ChronoGPT to GPT-2 XL, which has a comparable number of parameters. ChronoBERT's GLUE score stabilizes at 85, falling short of ModernBERT's 88. Likewise, ChronoGPT achieves a HellaSwag score of 37\% after 40 billion tokens, which is lower than the 48\% achieved by GPT-2 XL. We attribute this performance ceiling to the temporally-constrained nature of our training data, as continued training on the same dataset yields no further improvements.
% \footnote{The performance gap in the GLUE scores between ChronoBERT and the ModernBERT is likely stemming from both the computation and data limitations. The ModernBERT is trained on 1.8 trillion tokens with less repetition compared to ChronoBERT-1999 that is trained on 400 billion tokens with many repetitive runs.}
% \footnote{We obtained the Hellaswag evaluation score for the GPT‑2 model from \citet{karpathy_build_nanogpt_2024}, available at \url{https://github.com/karpathy/build-nanogpt/blob/master/play.ipynb}.}

\begin{figure}[!htb]
\begin{footnotesize}
\begin{center}
(a) ChronoBERT
\includegraphics[width=\linewidth]{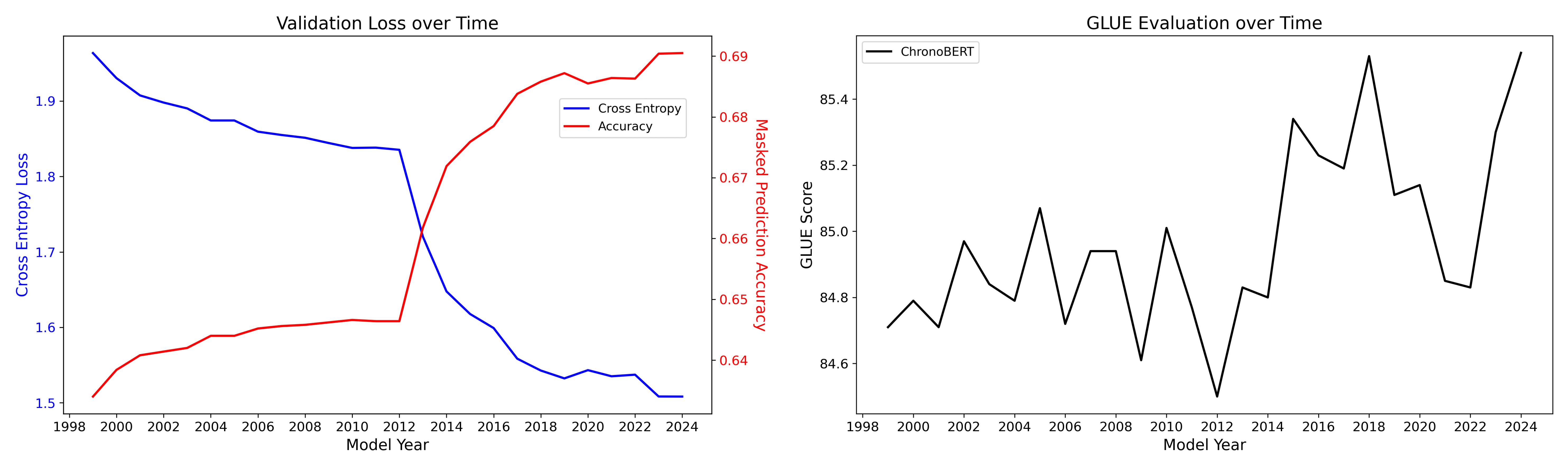}
(b) ChronoGPT
\includegraphics[width=\linewidth]{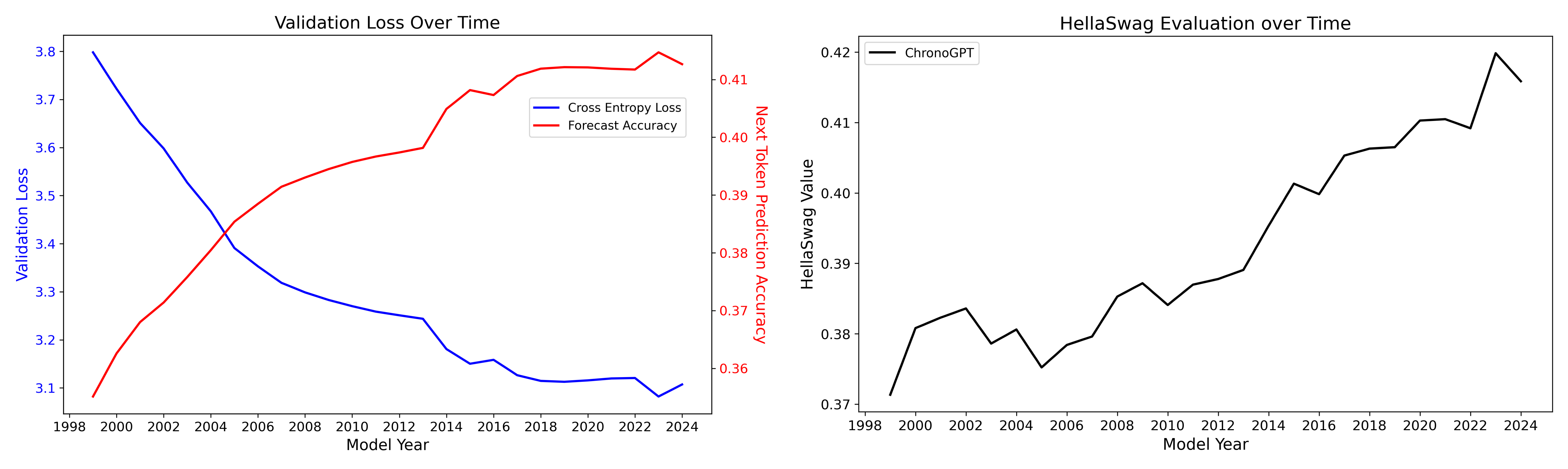}
\end{center}
\end{footnotesize}
\caption{Validation Loss and Evaluation Scores over Time}\label{fig:language_time}
\bigskip
\small
The left panel shows the validation loss, measured using cross-entropy loss and masked language prediction accuracy as the model is trained over time. The right panel displays the GLUE and HellaSwag scores as training progresses over time.
\end{figure}

Starting from these high-quality base models in the early period, we continue incremental pretraining using textual data afterward. We create model checkpoints for each year from 1999 to 2024 (26 models in total). Figure \ref{fig:language_time} presents our models' validation loss and language understanding scores as we train with incremental textual data over time.

We find that with the introduction of new data, the validation loss continues to drop. Starting in the year 2013, we introduce high-quality common crawl data. We witness a significant decrease in validation loss with the increase in data diversity. In the right panel of Figure \ref{fig:language_time}, the GLUE and HellaSwag scores for nearly all models exceed those of BERT and GPT-2, highlighting their superior language understanding and overall quality.\footnote{ModernBERT is trained on roughly 1.8 trillion tokens, over three times the 525 billion tokens used for $\text{ChronoBERT}_{2024}$. This much larger pre-training corpus largely explains its stronger GLUE performance.}

\subsection{Language Understanding}

We further compare language understanding against other models. Table \ref{tab:knowledge_cutoff} summarizes key characteristics of these models, including parameter counts, context lengths, and knowledge cutoffs.

The models evaluated include:

\textbf{$\text{ChronoBERT}_{t}$}: Our initial BERT-based model, $\text{ChronoBERT}_{1999}$ is pretrained on 460 billion tokens of pre-2000, diverse, high-quality, and open-source text data to ensure no leakage of data afterward.
Then, each year $t$ starting in 2000, we start from the model trained in the previous year and continue training it on data available in year $t$. Our final checkpoint of the ChronoBERT series, $\text{ChronoBERT}_{2024}$, is pretrained on 525 billion tokens of diverse, high-quality, and open-source text until December 2024.

\textbf{$\text{ChronoGPT}_{t}$}: Our initial GPT-based model, trained with the same chronologically consistent text data as ChronoBERT, ensures no lookahead bias by only utilizing information available at each point in time. The ChronoGPT series spans from 1999 to 2024, with each yearly checkpoint building upon the previous year’s model.

\textbf{Llama 3.1}: The 8B-parameter variant of the Llama 3.1 model by \citet{dubey2024llama}. \footnote{Model downloaded from \url{https://huggingface.co/meta-llama/Llama-3.1-8B}.}

\textbf{BERT}: The original BERT model trained on Wikipedia and the BookCorpus dataset by \citet{devlin-etal-2019-bert}.\footnote{Model downloaded from \url{https://huggingface.co/google-bert/bert-base-uncased}.}

\textbf{FinBERT}: A domain-specific model pretrained on financial texts, including regulatory filings, analyst reports, and earnings call transcripts, from \citet{huang2023finbert}.\footnote{Model downloaded from \url{https://huggingface.co/yiyanghkust/finbert-pretrain}.}

% \textbf{ModernBERT}: A modernized bidirectional encoder model with stronger downstream performance, higher speed, and higher memory efficiency, from \citet{warner2024smarter}\footnote{Model download from \url{https://github.com/AnswerDotAI/ModernBERT}.}

\textbf{StoriesLM}: A pretrained model from \citet{sarkar_storieslm_2024}, trained on historical news articles. We use the final version trained on data up to 1963.\footnote{Model downloaded from \url{https://huggingface.co/StoriesLM/StoriesLM-v1-1963}.}

% \textbf{GPT-2 xl}: The 1.5B-parameter version of OpenAI's GPT2 model from \citet{radford2019language}.\footnote{Model downloaded from \url{https://huggingface.co/openai-community/gpt2}.}

\begin{table}[!htb]
    \begin{center}
    \begin{tabularx}{\textwidth}{@{\hskip\tabcolsep\extracolsep\fill}l*{1}rrr}
    \toprule
         & Parameters &  Context Tokens \hspace{-0.15in} & Knowledge Cutoff \\
    \midrule
      $\text{ChronoBERT}_{1999}$ & 149M & 1,024 & December, 1999 \\
      \vdots & \vdots & \vdots & \vdots \\
      $\text{ChronoBERT}_{2024}$ & 149M & 1,024 & December, 2024 \\
      $\text{ChronoGPT}_{1999}$ & 1,552M & 1,792 & December, 1999 \\
      \vdots & \vdots & \vdots & \vdots \\
      $\text{ChronoGPT}_{2024}$ & 1,552M & 1,792 & December, 2024 \\
      Llama 3.1 & 8,030M & 128,000 & December 2023\\
      BERT & 110M & 512 & October, 2018 \\
      FinBERT & 110M & 512 & December, 2019 \\
      StoriesLM & 110M & 512 & December, 1963 \\
      % ModernBERT & 149M & 8,192 & December, 2024 \\
      % GPT-2 xl & 1,542M & 1,024 & Febrary, 2019 \\
    \bottomrule
    \end{tabularx}
    \end{center}
    \caption{Characteristics and Knowledge Cutoffs of Different LLMs}
    \label{tab:knowledge_cutoff}
    \bigskip
    \small
This table provides an overview of our chronologically consistent large language models (ChronoBERT and ChronoGPT) as well as several natural benchmark models, including their number of parameters, maximum context length, and knowledge cutoff dates.
\end{table}

\begin{table}[!htb]
    \begin{center}
    % \begin{footnotesize}
    % \setlength{\tabcolsep}{5pt}
    \begin{tabularx}{\textwidth}{l *{4}{Y}}
        \toprule
         & $\text{ChronoBERT}_{1999}$ & $\text{ChronoBERT}_{2024}$ & 
         $\text{ChronoGPT}_{1999}$ & $\text{ChronoGPT}_{2024}$ \\
        \cmidrule{2-5}
        COLA  & 57.32 & 56.32 & 35.50 & 48.20 \\
        SST2  & 91.82 & 92.58 & 90.60 & 92.40 \\
        MRPC  & 92.71 & 92.45 & 87.66 & 88.55 \\
        STSB  & 89.57 & 89.93 & 80.65 & 86.07 \\
        QQP   & 88.54 & 88.90 & 83.70 & 85.75 \\
        MNLI  & 86.19 & 86.89 & 82.27 & 84.91 \\
        QNLI  & 90.61 & 92.04 & 84.68 & 88.56 \\
        RTE   & 80.94 & 85.20 & 74.30 & 78.63 \\
        \cmidrule{2-5}
        GLUE  & 84.71 & 85.54 & 77.42 & 81.63 \\
        \midrule
        & Llama 3.1 & BERT & FinBERT & StoriesLM \\
        \cmidrule{2-5}
        COLA  & 55.86 & 57.59 & 28.99 & 46.85 \\
        SST2  & 95.49 & 92.62 & 89.03 & 90.44 \\
        MRPC  & 88.22 & 90.76 & 88.59 & 89.33 \\
        STSB  & 90.67 & 90.07 & 85.72 & 87.01 \\
        QQP   & 89.67 & 88.21 & 86.60 & 86.88 \\
        MNLI  & 89.59 & 84.98 & 79.23 & 79.78 \\
        QNLI  & 95.35 & 91.52 & 86.12 & 87.44 \\
        RTE   & 85.63 & 80.43 & 67.00 & 67.15 \\
        \cmidrule{2-5}
        GLUE  & 86.31 & 84.52 & 76.41 & 79.36 \\
        \bottomrule
    \end{tabularx}
%     \begin{tabularx}{\textwidth}{l *{8}{Y} c Y}
%     \toprule
%     & COLA & SST2 & MRPC & STSB & QQP & MNLI & QNLI & RTE & & GLUE \\
%     % \cmidrule{2-9} \cmidrule{11-11}
%     \midrule
%     $\text{ChronoBERT}_{1999}$ & 57.32 & 91.82 & 92.71 & 89.57 & 88.54 & 86.19 & 90.61 & 80.94 & & 84.71 \\
%     $\text{ChronoBERT}_{2024}$ & 56.32 & 92.58 & 92.45 & 89.93 & 88.90 & 86.89 & 92.04 & 85.20 & & 85.54 \\
%     $\text{ChronoGPT}_{1999}$  & 37.13 & 89.68 & 82.92 & 81.57 & 82.43 & 77.63 & 84.94 & 67.08 & & 75.42 \\
%     $\text{ChronoGPT}_{2024}$  & 31.70 & 88.53 & 85.34 & 82.58 & 83.53 & 79.15 & 85.98 & 67.80 & & 75.58 \\
%     BERT      & 57.59 & 92.62 & 90.76 & 90.07 & 88.21 & 84.98 & 91.52 & 80.43 & & 84.52 \\
%     FinBERT   & 28.99 & 89.03 & 88.59 & 85.72 & 86.60 & 79.23 & 86.12 & 67.00 & & 76.41 \\
%     StoriesLM & 46.85 & 90.44 & 89.33 & 87.01 & 86.88 & 79.78 & 87.44 & 67.15 & & 79.36 \\
%     Llama 3.1 & 55.86 & 95.49 & 88.22 & 90.67 & 89.67 & 89.59 & 95.35 & 85.63 & & 86.31 \\
%     \bottomrule
% \end{tabularx}
    % \end{footnotesize}
    \end{center}
    \caption{GLUE Score Evaluations for Different LLMs}\label{tab:glue_scores}
    \bigskip
    \small
This table compares the GLUE benchmark scores of our chronologically consistent large language models (ChronoBERT and ChronoGPT) as well as several natural benchmark models. Tasks are grouped into three categories: (1) Single-sentence classification (COLA, SST2), (2) Paraphrase/semantic similarity (MRPC, STSB, QQP), and (3) Natural language inference (MNLI, QNLI, RTE). The final row shows the average GLUE score across all tasks.
\end{table}

Table \ref{tab:glue_scores} displays the GLUE scores for eight different models. While Llama 3.1 achieves the highest overall GLUE score (86.31), $\text{ChronoBERT}_{2024}$ and $\text{ChronoBERT}_{1999}$, standing at less than 1/50 of Llama's size, still deliver strong performance, surpassing BERT (84.52) while substantially outperforming StoriesLM (79.36) and FinBERT (76.41). Notably, both ChronoBERT models Pareto dominate the last two models—they outperform StoriesLM (designed to avoid lookahead bias) and FinBERT (domain-specific) across all individual tasks, with particularly large margins on COLA, RTE, and MNLI. This performance advantage persists even in BERT-competitive tasks like MRPC and QQP.

Similarly, the ChronoGPT models showcase the effectiveness of our training methodology, with $\text{ChronoGPT}_{2024}$ (81.63) significantly outperforming FinBERT and StoriesLM. The most salient finding for ChronoGPT is its growth; it gains over 4 points on the GLUE benchmark from its 1999 counterpart (77.42), with especially large improvements in grammatical correctness (COLA) and inference (RTE). This confirms that chronological data enrichment progressively enhances language understanding in both encoder-only and decoder-only architectures.

% By contrast, $\text{ChronoGPT}_{2024}$ and $\text{ChronoGPT}_{1999}$ do not match ChronoBERT’s scores on GLUE. Their lower performance reflects the nature of an autoregressive (decoder-only) architecture, which processes text strictly left to right for text generation. While this approach can excel at generative tasks, it often underperforms on classification benchmarks that benefit from bidirectional processing of the entire input.

Overall, ChronoBERT and ChronoGPT exhibit strong language understanding, with ChronoBERT coming remarkably close to Llama 3.1's performance despite Llama 3.1 being a much larger and more recent model. The performance gap between ChronoBERT and StoriesLM likely stems from differences in training scale (460B versus 19B tokens) and the quality and diversity of the training data (ChronoBERT's high-quality corpus versus StoriesLM's unfiltered news-only dataset). Similarly, ChronoBERT and ChronoGPT's significant edge over FinBERT highlights the importance of diverse and high-quality pretraining data, as FinBERT's domain-specific financial texts lack comprehensive quality checks.

Importantly, our chronologically consistent models prove that lookahead bias can be eliminated without compromising the core capabilities expected of modern LLMs. This makes them particularly valuable for applications that require good language understanding and no lookahead bias. By maintaining both strong language understanding and temporal integrity, our models are particularly well-suited for applications where preserving the timeline of information is critical, making them a robust choice for a wide range of downstream tasks.

\subsection{Validation of Chronological Consistency}

Although we exercise great caution in curating our training data to ensure they include only information that we believe was available by a specific date, this process is not necessarily without imperfections. Potential inaccuracies in recorded publication dates---for example, if a hard copy document was digitized via optical character recognition (OCR) and an incorrect date was captured in the process---may result in the unintended inclusion of information that was not actually available at the intended time, thereby introducing traces of lookahead bias into our models.

To detect leakage in the textual data used to pretrain our chronologically consistent models, we evaluate them on events occurring after the model’s knowledge cutoff. This involves constructing a textual sequence with a missing token representing time-specific information beyond the model’s cutoff. We use each model to tokenize the sentence and predict the missing token. Due to the bi-directional nature of ChronoBERT, the missing token is a masked token in the middle of a sentence, with the surrounding language giving context for the prediction. Consider the following sentence involving an incoming U.S. president, where the president's name is masked and is the target of the prediction:

\begin{table}[!htb]
    \resizebox{\textwidth}{!}{
    \begin{tabular}{lccccccccc}
        % \multicolumn{10}{l}{Panel A: Input prompt} \\
        % \toprule
        %  \multicolumn{10}{p{1.25\textwidth}}{After the \{$\text{year}-1$\} U.S. presidential election, President [MASK] was inaugurated as U.S. President in the year \{year\}.}\\
        % \bottomrule
        % \\
        % \multicolumn{10}{l}{Panel B: Output predictions} \\
        \toprule
        & \multicolumn{6}{c}{Prompt year} & & \multicolumn{2}{c}{Accuracy}\\
        \cmidrule{2-7} \cmidrule{9-10}
         & 1993 & 2001 & 2009 & 2017 & 2021 & 2025 & & Pre-cutoff & Post-cutoff \\
        \midrule
        Correct output & Clinton & Bush & Obama & Trump & Biden & Trump \\
        \midrule
        BERT              & \textcolor{blue}{\textbf{Clinton}}  & Clinton  & \textcolor{blue}{\textbf{Obama}}  & Obama  &\cellcolor{lightgray}Obama  &\cellcolor{lightgray}Obama & & $2/4$ & \cellcolor{lightgray}$0/2$\\
        ModernBERT              & \textcolor{blue}{\textbf{Clinton}}  & \textcolor{blue}{\textbf{Bush}}  & \textcolor{blue}{\textbf{Obama}}  & \textcolor{blue}{\textbf{Trump}}  & \textcolor{blue}{\textbf{Biden}}  & \cellcolor{lightgray}\textcolor{blue}{\textbf{Trump}} & & $5/5$ & \cellcolor{lightgray}$1/1$\\
        \midrule
        $\text{ChronoBERT}_{1999}$   &  Roosevelt  & \cellcolor{lightgray}Hoover  & \cellcolor{lightgray}Roosevelt  & \cellcolor{lightgray}Hoover  & \cellcolor{lightgray}Hoover  & \cellcolor{lightgray}Washington & &  $0/1$ & \cellcolor{lightgray} $0/5$\\
        $\text{ChronoBERT}_{2000}$   & \textcolor{blue}{\textbf{Clinton}}  & \cellcolor{lightgray}Clinton  & \cellcolor{lightgray}Clinton  & \cellcolor{lightgray}Clinton  & \cellcolor{lightgray}Clinton  & \cellcolor{lightgray}Wilson &  & $1/1$ & \cellcolor{lightgray} $0/5$\\
        $\text{ChronoBERT}_{2001}$   & \textcolor{blue}{\textbf{Clinton}}  & Clinton  & \cellcolor{lightgray}Clinton  & \cellcolor{lightgray}Clinton  & \cellcolor{lightgray}Clinton  & \cellcolor{lightgray}Washington & &  $1/2$ & \cellcolor{lightgray} $0/4$\\
        $\text{ChronoBERT}_{2002}$  & \textcolor{blue}{\textbf{Clinton}}  & \textcolor{blue}{\textbf{Bush}}  & \cellcolor{lightgray}Bush  & \cellcolor{lightgray}Clinton  & \cellcolor{lightgray}Bush  & \cellcolor{lightgray}Clinton & &  $2/2$ & \cellcolor{lightgray} $0/4$\\
        $\text{ChronoBERT}_{2003}$  & \textcolor{blue}{\textbf{Clinton}}  & \textcolor{blue}{\textbf{Bush}} & \cellcolor{lightgray}Bush  & \cellcolor{lightgray}Clinton  & \cellcolor{lightgray}Bush & \cellcolor{lightgray}Clinton & &  $2/2$ & \cellcolor{lightgray} $0/4$\\
        $\text{ChronoBERT}_{2004}$  & \textcolor{blue}{\textbf{Clinton}}  & \textcolor{blue}{\textbf{Bush}}  & \cellcolor{lightgray}Bush  & \cellcolor{lightgray}Clinton  & \cellcolor{lightgray}Bush  & \cellcolor{lightgray}Clinton & &  $2/2$ & \cellcolor{lightgray} $0/4$\\
        $\text{ChronoBERT}_{2005}$  & \textcolor{blue}{\textbf{Clinton}}  & \textcolor{blue}{\textbf{Bush}}  & \cellcolor{lightgray}Bush  & \cellcolor{lightgray}Bush & \cellcolor{lightgray}Bush  & \cellcolor{lightgray}Clinton & &  $2/2$ & \cellcolor{lightgray} $0/4$\\
        $\text{ChronoBERT}_{2006}$  & \textcolor{blue}{\textbf{Clinton}}  & \textcolor{blue}{\textbf{Bush}}  & \cellcolor{lightgray}Bush  & \cellcolor{lightgray}Bush  & \cellcolor{lightgray}Bush  & \cellcolor{lightgray}Clinton & &  $2/2$ & \cellcolor{lightgray} $0/4$\\
        $\text{ChronoBERT}_{2007}$  & \textcolor{blue}{\textbf{Clinton}}  & \textcolor{blue}{\textbf{Bush}}  & \cellcolor{lightgray}Bush  & \cellcolor{lightgray}Bush  & \cellcolor{lightgray}Bush  & \cellcolor{lightgray}Monroe & &  $2/2$ & \cellcolor{lightgray} $0/4$\\
        $\text{ChronoBERT}_{2008}$  & \textcolor{blue}{\textbf{Clinton}}  & \textcolor{blue}{\textbf{Bush}}  &\cellcolor{lightgray}Bush  &\cellcolor{lightgray}Obama  &\cellcolor{lightgray}Bush  &\cellcolor{lightgray}Wilson & &  $2/2$ & \cellcolor{lightgray} $0/4$\\
        $\text{ChronoBERT}_{2009}$  & \textcolor{blue}{\textbf{Clinton}}  & Clinton & \textcolor{blue}{\textbf{Obama}}  &\cellcolor{lightgray}Obama  &\cellcolor{lightgray}Obama  &\cellcolor{lightgray}Wilson &  &  $2/3$ & \cellcolor{lightgray} $0/3$\\
        $\text{ChronoBERT}_{2010}$  & \textcolor{blue}{\textbf{Clinton}}  & Obama & \textcolor{blue}{\textbf{Obama}}  &\cellcolor{lightgray}Obama  &\cellcolor{lightgray}Obama  &\cellcolor{lightgray}Wilson & &  $2/3$ & \cellcolor{lightgray} $0/3$\\
        $\text{ChronoBERT}_{2011}$  & \textcolor{blue}{\textbf{Clinton}}  & Clinton  & \textcolor{blue}{\textbf{Obama}}  &\cellcolor{lightgray}Obama  &\cellcolor{lightgray}Obama  &\cellcolor{lightgray}Wilson & &  $2/3$ & \cellcolor{lightgray} $0/3$\\
        $\text{ChronoBERT}_{2012}$  & Obama  & Obama  & \textcolor{blue}{\textbf{Obama}}  &\cellcolor{lightgray}Obama  &\cellcolor{lightgray}Obama  &\cellcolor{lightgray}Obama & &  $1/3$ & \cellcolor{lightgray} $0/3$\\
        $\text{ChronoBERT}_{2013}$  & \textcolor{blue}{\textbf{Clinton}}  & Obama  & \textcolor{blue}{\textbf{Obama}}  &\cellcolor{lightgray}Obama  &\cellcolor{lightgray}Obama  &\cellcolor{lightgray}Monroe & &  $2/3$ & \cellcolor{lightgray} $0/3$\\
        $\text{ChronoBERT}_{2014}$  & \textcolor{blue}{\textbf{Clinton}}  & \textcolor{blue}{\textbf{Bush}}  & \textcolor{blue}{\textbf{Obama}}  &\cellcolor{lightgray}Obama  &\cellcolor{lightgray}Obama  &\cellcolor{lightgray}Monroe & &  $3/3$ & \cellcolor{lightgray} $0/3$\\
        $\text{ChronoBERT}_{2015}$ & \textcolor{blue}{\textbf{Clinton}}  & Clinton  & \textcolor{blue}{\textbf{Obama}}  &\cellcolor{lightgray}Obama  &\cellcolor{lightgray}Obama  &\cellcolor{lightgray}Monroe & &  $2/3$ & \cellcolor{lightgray} $0/3$\\
        $\text{ChronoBERT}_{2016}$  & \textcolor{blue}{\textbf{Clinton}}  & \textcolor{blue}{\textbf{Bush}}  & \textcolor{blue}{\textbf{Obama}}  &\cellcolor{lightgray}Obama  &\cellcolor{lightgray}Obama  &\cellcolor{lightgray}Obama & &  $3/3$ & \cellcolor{lightgray} $0/3$\\
        $\text{ChronoBERT}_{2017}$  & \textcolor{blue}{\textbf{Clinton}}  & \textcolor{blue}{\textbf{Bush}}  & \textcolor{blue}{\textbf{Obama}}  & \textcolor{blue}{\textbf{Trump}}  &\cellcolor{lightgray}Trump  &\cellcolor{lightgray}Monroe & &  $4/4$ & \cellcolor{lightgray} $0/2$\\
        $\text{ChronoBERT}_{2018}$  & \textcolor{blue}{\textbf{Clinton}}  & \textcolor{blue}{\textbf{Bush}}  & \textcolor{blue}{\textbf{Obama}}  & \textcolor{blue}{\textbf{Trump}}  &\cellcolor{lightgray}Trump  &\cellcolor{lightgray}Obama & &  $4/4$ & \cellcolor{lightgray} $0/2$\\
        $\text{ChronoBERT}_{2019}$  & \textcolor{blue}{\textbf{Clinton}}  & \textcolor{blue}{\textbf{Bush}}  & \textcolor{blue}{\textbf{Obama}}  & \textcolor{blue}{\textbf{Trump}}  &\cellcolor{lightgray}Trump  &\cellcolor{lightgray}Obama & &  $4/4$ & \cellcolor{lightgray} $0/2$\\
        $\text{ChronoBERT}_{2020}$  & \textcolor{blue}{\textbf{Clinton}}  & \textcolor{blue}{\textbf{Bush}}  & \textcolor{blue}{\textbf{Obama}}  & \textcolor{blue}{\textbf{Trump}}  &\cellcolor{lightgray}Trump  &\cellcolor{lightgray}\textcolor{blue}{\textbf{Trump}} & & $4/4$ & \cellcolor{lightgray} $1/2$\\
        $\text{ChronoBERT}_{2021}$  & \textcolor{blue}{\textbf{Clinton}}  & Clinton  & \textcolor{blue}{\textbf{Obama}}  & \textcolor{blue}{\textbf{Trump}}  & \textcolor{blue}{\textbf{Biden}}  &\cellcolor{lightgray}Biden & &  $4/5$ & \cellcolor{lightgray} $0/1$\\
        $\text{ChronoBERT}_{2022}$  & \textcolor{blue}{\textbf{Clinton}}  & \textcolor{blue}{\textbf{Bush}}  & \textcolor{blue}{\textbf{Obama}}  & \textcolor{blue}{\textbf{Trump}}  & \textcolor{blue}{\textbf{Biden}}  &\cellcolor{lightgray}Biden & &  $5/5$ & \cellcolor{lightgray} $0/1$\\
        $\text{ChronoBERT}_{2023}$  & \textcolor{blue}{\textbf{Clinton}}  & \textcolor{blue}{\textbf{Bush}}  & \textcolor{blue}{\textbf{Obama}}  & \textcolor{blue}{\textbf{Trump}}  & \textcolor{blue}{\textbf{Biden}}  &\cellcolor{lightgray}Biden & &  $5/5$ & \cellcolor{lightgray} $0/1$\\
        $\text{ChronoBERT}_{2024}$  & \textcolor{blue}{\textbf{Clinton}}  & \textcolor{blue}{\textbf{Bush}}  & \textcolor{blue}{\textbf{Obama}}  & \textcolor{blue}{\textbf{Trump}}  & \textcolor{blue}{\textbf{Biden}}  &\cellcolor{lightgray}Biden & &  $5/5$ & \cellcolor{lightgray} $0/1$\\
        \hline
        \multicolumn{8}{l}{$\text{ChronoBERT}_{1999}$ through $\text{ChronoBERT}_{2024}$} & $68/78$ & \cellcolor{lightgray}$1/78$\\
        \bottomrule
    \end{tabular}
    }
    \caption{Masked-Token Predictions of U.S. Presidents using ChronoBERT}
    \label{tab:presidents_chronobert}
        \bigskip
    \footnotesize
This table displays ChronoBERT's masked-token predictions in a sentence describing an incoming U.S. president, where the president's name is the target of the prediction. The input prompt is, 
\begin{quotation}
  ``After the \{$\text{year}-1$\} U.S. presidential election, President [MASK] was inaugurated as U.S. President in the year \{year\}."   
\end{quotation}
For each masked token, the model's prediction is the token with the highest probability, chosen deterministically. The gray-shaded area denotes predictions covering the post-knowledge cutoff period. The blue text highlights correct predictions. For comparison, predictions from BERT (released in 2018) and ModernBERT (released in 2024) are also included.
\end{table}

\begin{table}[!htb]
    \resizebox{\textwidth}{!}{
    \begin{tabular}{lccccccccc}
        \toprule
        & \multicolumn{6}{c}{Prompt year} & & \multicolumn{2}{c}{Accuracy}\\
        \cmidrule{2-7} \cmidrule{9-10}
         & 1993 & 2001 & 2009 & 2017 & 2021 & 2025 & & Pre-cutoff & Post-cutoff \\
        \midrule
        Correct output & Bill Clinton & George W. (Bush) & Barack Obama & Donald Trump & Joe Biden & Donald Trump\\
        \midrule
        GPT-2              & \textcolor{blue}{\textbf{Bill Clinton}}  & Bill Clinton  & \textcolor{blue}{\textbf{Barack Obama}}  & \textcolor{blue}{\textbf{Donald Trump}}  &\cellcolor{lightgray}George W.  &\cellcolor{lightgray}George W. & & $3/4$ & \cellcolor{lightgray}$0/2$\\
        GPT-2 XL             & \textcolor{blue}{\textbf{Bill Clinton}}  & \textcolor{blue}{\textbf{George W.}}  & \textcolor{blue}{\textbf{Barack Obama}}  & \textcolor{blue}{\textbf{Donald Trump}}  &\cellcolor{lightgray}James A.  &\cellcolor{lightgray}James Mattis & & $4/4$ & \cellcolor{lightgray}$0/2$\\
        \midrule
        $\text{ChronoGPT}_{1999}$   &  \textcolor{blue}{\textbf{Bill Clinton}}  & \cellcolor{lightgray}Bill Clinton  & \cellcolor{lightgray}Bill Clinton  & \cellcolor{lightgray}Obama\textbackslash n  & \cellcolor{lightgray}Obama\textbackslash n  & \cellcolor{lightgray}Obama\textbackslash n & &  $1/1$ & \cellcolor{lightgray} $0/5$\\
        $\text{ChronoGPT}_{2000}$   & \textcolor{blue}{\textbf{Bill Clinton}}  & \cellcolor{lightgray}Bill Clinton  & \cellcolor{lightgray}Bill Clinton  & \cellcolor{lightgray}Bill Clinton  & \cellcolor{lightgray}Bill Clinton  & \cellcolor{lightgray}Bill Clinton &  & $1/1$ & \cellcolor{lightgray} $0/5$\\
        $\text{ChronoGPT}_{2001}$   & George W.  & \textcolor{blue}{\textbf{George W.}}  & \cellcolor{lightgray}Bill Clinton  & \cellcolor{lightgray}George W.  & \cellcolor{lightgray}George W.  & \cellcolor{lightgray}George W. & &  $1/2$ & \cellcolor{lightgray} $0/4$\\
        $\text{ChronoGPT}_{2002}$  & \textcolor{blue}{\textbf{Bill Clinton}}  & \textcolor{blue}{\textbf{George W.}}  & \cellcolor{lightgray}George W.  & \cellcolor{lightgray}George W. & \cellcolor{lightgray}George W.  & \cellcolor{lightgray}George W. & &  $2/2$ & \cellcolor{lightgray} $0/4$\\
        $\text{ChronoGPT}_{2003}$  & \textcolor{blue}{\textbf{Bill Clinton}}  & \textcolor{blue}{\textbf{George W.}}  & \cellcolor{lightgray}George W.  & \cellcolor{lightgray}George W. & \cellcolor{lightgray}George W.  & \cellcolor{lightgray}George W. & &  $2/2$ & \cellcolor{lightgray} $0/4$\\
        $\text{ChronoGPT}_{2004}$  & \textcolor{blue}{\textbf{Bill Clinton}}  & \textcolor{blue}{\textbf{George W.}}  & \cellcolor{lightgray}George W.  & \cellcolor{lightgray}George W. & \cellcolor{lightgray}George W.  & \cellcolor{lightgray}George W. & &  $2/2$ & \cellcolor{lightgray} $0/4$\\
        $\text{ChronoGPT}_{2005}$  & \textcolor{blue}{\textbf{Bill Clinton}}  & \textcolor{blue}{\textbf{George W.}}  & \cellcolor{lightgray}George W.  & \cellcolor{lightgray}George W. & \cellcolor{lightgray}George W.  & \cellcolor{lightgray}George W. & &  $2/2$ & \cellcolor{lightgray} $0/4$\\
        $\text{ChronoGPT}_{2006}$  & \textcolor{blue}{\textbf{Bill Clinton}}  & \textcolor{blue}{\textbf{George W.}}  & \cellcolor{lightgray}George W.  & \cellcolor{lightgray}George W. & \cellcolor{lightgray}George W.  & \cellcolor{lightgray}George W. & &  $2/2$ & \cellcolor{lightgray} $0/4$\\
        $\text{ChronoGPT}_{2007}$  & \textcolor{blue}{\textbf{Bill Clinton}}  & \textcolor{blue}{\textbf{George W.}}  & \cellcolor{lightgray}George W.  & \cellcolor{lightgray}George W. & \cellcolor{lightgray}George W.  & \cellcolor{lightgray}George W. & &  $2/2$ & \cellcolor{lightgray} $0/4$\\
        $\text{ChronoGPT}_{2008}$  & \textcolor{blue}{\textbf{Bill Clinton}}  & \textcolor{blue}{\textbf{George W.}}  & \cellcolor{lightgray}George W.  & \cellcolor{lightgray}George W. & \cellcolor{lightgray}George W.  & \cellcolor{lightgray}George W. & &  $2/2$ & \cellcolor{lightgray} $0/4$\\
        $\text{ChronoGPT}_{2009}$  & \textcolor{blue}{\textbf{Bill Clinton}}  & \textcolor{blue}{\textbf{George W.}} & \textcolor{blue}{\textbf{Barack Obama}}  &\cellcolor{lightgray}Barack Obama  &\cellcolor{lightgray}Barack Obama  &\cellcolor{lightgray}Barack Obama &  &  $3/3$ & \cellcolor{lightgray} $0/3$\\
        $\text{ChronoGPT}_{2010}$  & \textcolor{blue}{\textbf{Bill Clinton}}  & \textcolor{blue}{\textbf{George W.}} & \textcolor{blue}{\textbf{Barack Obama}}  &\cellcolor{lightgray}Barack Obama  &\cellcolor{lightgray}Barack Obama  &\cellcolor{lightgray}Barack Obama &  &  $3/3$ & \cellcolor{lightgray} $0/3$\\
        $\text{ChronoGPT}_{2011}$  & \textcolor{blue}{\textbf{Bill Clinton}}  & \textcolor{blue}{\textbf{George W.}} & \textcolor{blue}{\textbf{Barack Obama}}  &\cellcolor{lightgray}Barack Obama  &\cellcolor{lightgray}Barack Obama  &\cellcolor{lightgray}Barack Obama &  &  $3/3$ & \cellcolor{lightgray} $0/3$\\
        $\text{ChronoGPT}_{2012}$  & \textcolor{blue}{\textbf{Bill Clinton}}  & \textcolor{blue}{\textbf{George W.}} & \textcolor{blue}{\textbf{Barack Obama}}  &\cellcolor{lightgray}Barack Obama  &\cellcolor{lightgray}Barack Obama  &\cellcolor{lightgray}Barack Obama &  &  $3/3$ & \cellcolor{lightgray} $0/3$\\
        $\text{ChronoGPT}_{2013}$  & \textcolor{blue}{\textbf{Bill Clinton}}  & \textcolor{blue}{\textbf{George W.}} & \textcolor{blue}{\textbf{Barack Obama}}  &\cellcolor{lightgray}Barack Obama  &\cellcolor{lightgray}Barack Obama  &\cellcolor{lightgray}Barack Obama &  &  $3/3$ & \cellcolor{lightgray} $0/3$\\
        $\text{ChronoGPT}_{2014}$  & \textcolor{blue}{\textbf{Bill Clinton}}  & \textcolor{blue}{\textbf{George W.}} & \textcolor{blue}{\textbf{Barack Obama}}  &\cellcolor{lightgray}Barack Obama  &\cellcolor{lightgray}Barack Obama  &\cellcolor{lightgray}Barack Obama &  &  $3/3$ & \cellcolor{lightgray} $0/3$\\
        $\text{ChronoGPT}_{2015}$  & \textcolor{blue}{\textbf{Bill Clinton}}  & \textcolor{blue}{\textbf{George W.}} & \textcolor{blue}{\textbf{Barack Obama}}  &\cellcolor{lightgray}Barack Obama  &\cellcolor{lightgray}Barack Obama  &\cellcolor{lightgray}Barack Obama &  &  $3/3$ & \cellcolor{lightgray} $0/3$\\
        $\text{ChronoGPT}_{2016}$  & \textcolor{blue}{\textbf{Bill Clinton}}  & \textcolor{blue}{\textbf{George W.}} & \textcolor{blue}{\textbf{Barack Obama}}  &\cellcolor{lightgray}Barack Obama  &\cellcolor{lightgray}Barack Obama  &\cellcolor{lightgray}Barack Obama &  &  $3/3$ & \cellcolor{lightgray} $0/3$\\
        $\text{ChronoGPT}_{2017}$  & \textcolor{blue}{\textbf{Bill Clinton}}  & \textcolor{blue}{\textbf{George W.}} & \textcolor{blue}{\textbf{Barack Obama}}  & \textcolor{blue}{\textbf{Donald Trump}}   &\cellcolor{lightgray}Donald Trump &\cellcolor{lightgray}\textcolor{blue}{\textbf{Donald Trump}} &  &  $4/4$ & \cellcolor{lightgray} $1/2$\\
        $\text{ChronoGPT}_{2018}$  & \textcolor{blue}{\textbf{Bill Clinton}}  & \textcolor{blue}{\textbf{George W.}} & \textcolor{blue}{\textbf{Barack Obama}}  & \textcolor{blue}{\textbf{Donald Trump}}   &\cellcolor{lightgray}Donald Trump &\cellcolor{lightgray}\textcolor{blue}{\textbf{Donald Trump}} &  &  $4/4$ & \cellcolor{lightgray} $1/2$\\
        $\text{ChronoGPT}_{2019}$  & \textcolor{blue}{\textbf{Bill Clinton}}  & \textcolor{blue}{\textbf{George W.}} & \textcolor{blue}{\textbf{Barack Obama}}  & \textcolor{blue}{\textbf{Donald Trump}}   &\cellcolor{lightgray}Donald Trump &\cellcolor{lightgray}\textcolor{blue}{\textbf{Donald Trump}} &  &  $4/4$ & \cellcolor{lightgray} $1/2$\\
        $\text{ChronoGPT}_{2020}$  & \textcolor{blue}{\textbf{Bill Clinton}}  & \textcolor{blue}{\textbf{George W.}} & \textcolor{blue}{\textbf{Barack Obama}}  & \textcolor{blue}{\textbf{Donald Trump}}   &\cellcolor{lightgray}Donald Trump &\cellcolor{lightgray}\textcolor{blue}{\textbf{Donald Trump}} &  &  $4/4$ & \cellcolor{lightgray} $1/2$\\
        $\text{ChronoGPT}_{2021}$  & \textcolor{blue}{\textbf{Bill Clinton}}  & \textcolor{blue}{\textbf{George W.}} & \textcolor{blue}{\textbf{Barack Obama}}  & \textcolor{blue}{\textbf{Donald Trump}}   & \textcolor{blue}{\textbf{Joe Biden}}  &\cellcolor{lightgray} Joe Biden &  &  $5/5$ & \cellcolor{lightgray} $0/1$\\
        $\text{ChronoGPT}_{2022}$  & \textcolor{blue}{\textbf{Bill Clinton}}  & \textcolor{blue}{\textbf{George W.}} & \textcolor{blue}{\textbf{Barack Obama}}  & \textcolor{blue}{\textbf{Donald Trump}}   & \textcolor{blue}{\textbf{Joe Biden}}  &\cellcolor{lightgray} Joe Biden &  &  $5/5$ & \cellcolor{lightgray} $0/1$\\
        $\text{ChronoGPT}_{2023}$  & \textcolor{blue}{\textbf{Bill Clinton}}  & \textcolor{blue}{\textbf{George W.}} & \textcolor{blue}{\textbf{Barack Obama}}  & \textcolor{blue}{\textbf{Donald Trump}}   & \textcolor{blue}{\textbf{Joe Biden}}  &\cellcolor{lightgray} Joe Biden &  &  $5/5$ & \cellcolor{lightgray} $0/1$\\
        $\text{ChronoGPT}_{2024}$  & \textcolor{blue}{\textbf{Bill Clinton}}  & \textcolor{blue}{\textbf{George W.}} & \textcolor{blue}{\textbf{Barack Obama}}  & \textcolor{blue}{\textbf{Donald Trump}}   & \textcolor{blue}{\textbf{Joe Biden}}  &\cellcolor{lightgray} Joe Biden &  &  $5/5$ & \cellcolor{lightgray} $0/1$\\
        \hline
        \multicolumn{8}{l}{$\text{ChronoGPT}_{1999}$ through $\text{ChronoGPT}_{2024}$} & $77/78$ & \cellcolor{lightgray}$4/78$\\
        \bottomrule
    \end{tabular}
    }
    \caption{Next-Token Predictions of U.S. Presidents using ChronoGPT}
    \label{tab:presidents_chronogpt}
        \bigskip
    \footnotesize
This table displays ChronoGPT's next-token predictions for a list including the incoming U.S. president and his three most recent predecessors, where the name of the most recent president is the target of the prediction. The input prompt is, 
\begin{quotation}
``U.S. Presidents in chronological order:

Took office in \{year$_{p-3}$\}: President \{name$_{p-3}$\}

Took office in \{year$_{p-2}$\}: President \{name$_{p-2}$\}

Took office in \{year$_{p-1}$\}: President \{name$_{p-1}$\}

Took office in \{year$_{p}$\}: President \rule{2cm}{0.1mm}". 
\end{quotation}
Each sequence includes exactly two predicted tokens, with the most probable token selected deterministically at each step. The gray-shaded area denotes predictions covering the post-knowledge cutoff period, including election years where the president-elect has yet to be inaugurated. The blue text highlights correct predictions. For comparison, predictions from GPT-2 and GPT-2 XL (released in 2019) are also included.
\end{table}
% \textbf{``U.S. Presidents in chronological order:}

% \textbf{Took office in \{year$_{\pi-3}$\}: President \{name$_{\pi-3}$\}}

% \textbf{Took office in \{year$_{\pi-2}$\}: President \{name$_{\pi-2}$\}}

% \textbf{Took office in \{year$_{\pi-1}$\}: President \{name$_{\pi-1}$\}}

% \textbf{Took office in \{year$_{\pi}$\}: President"}.
\begin{quotation}
``After the \{$\text{year}-1$\} U.S. presidential election, President [MASK] was inaugurated as U.S. President in the year \{year\}."
\end{quotation}
As for ChronoGPT, due to its autoregressive nature, the missing token is at the end of a textual sequence, with the preceding context guiding the prediction. We construct a textual sequence that is essentially a few-shot learning exercise. Denoting the most recent president as the $p^{\text{th}}$ president, consider the following list involving the incoming U.S. president and his three most recent predecessors, where the name of the most recent president (name$_p$) is omitted and is the target of the prediction:

\begin{quotation}
``U.S. Presidents in chronological order:

Took office in \{year$_{p-3}$\}: President \{name$_{p-3}$\}

Took office in \{year$_{p-2}$\}: President \{name$_{p-2}$\}

Took office in \{year$_{p-1}$\}: President \{name$_{p-1}$\}

Took office in \{year$_{p}$\}: President \rule{2cm}{0.1mm}".
\end{quotation}
Table \ref{tab:presidents_chronobert} presents the masked-token predictions using ChronoBERT, and Table \ref{tab:presidents_chronogpt} presents the next-token predictions using ChronoGPT. In both tables, the gray-shaded area in the top-right indicates predictions in the post-knowledge cutoff period, and the non-shaded lower-right denotes predictions strictly in the pre-knowledge cutoff period. Correct predictions are highlighted in blue. 

If a model is well-trained with high-quality and time-relevant textual data, it should be able to identify presidents taking office before its knowledge cutoff with a high degree of accuracy. At the same time, an absence of leakage would be evidenced by the model’s inability to predict any president elected for the first time after that cutoff, confirming that no future information was inadvertently included in the training data. In other words, for a well-trained model that is perfectly chronologically consistent, we would expect most predictions in the non-shaded area to appear in blue text, while all predictions in the gray-shaded area should appear in black text.

Here are the results of the tests involving U.S. presidents. In the pre-cutoff period represented by the non-shaded area, ChronoBERT correctly makes a majority of predictions (68 out of 78), outperforming the original BERT model (2 out of 4), though falling short of ModernBERT's perfect performance. ChronoGPT also correctly makes most predictions (77 out of 78), outperforming GPT-2 (3 out of 4), and coming very close to GPT-2 XL's perfect performance. In contrast, during the post-cutoff period represented by the gray-shaded area, none of the ChronoBERT or ChronoGPT models correctly predict a future president in his first term. 
Notably, one ChronoBERT model ($\text{ChronoBERT}_{2020}$) and four ChronoGPT models ($\text{ChronoGPT}_{2017}$ through $\text{ChronoGPT}_{2020}$) correctly predict President Trump’s second non-consecutive term in 2025---this reflects a tendency to favor either a recently elected or historically notable president in an out-of-knowledge context. For ChronoGPT in particular, this also reflects a tendency toward autoregressive repetition, as President Trump is the latest president provided in the input sequence. Therefore, none of the ChronoBERT or ChronoGPT models demonstrates foreknowledge of a future first-term president.
% The sole correct prediction is for $\text{ChronoBERT}_{2020}$, which suggests President Trump’s second non-consecutive term after the 2024 election; this reflects a tendency to favor either a recently elected or historically notable president in an out-of-knowledge context. Therefore, none of the ChronoBERT models has foreknowledge of any future first-term president.

\begin{table}[!htb]
    \resizebox{\textwidth}{!}{
    \begin{tabular}{lP{2.7cm}P{2.7cm}P{2.7cm}P{2.7cm}P{2.7cm}cccc}
        \multicolumn{10}{l}{Panel A: Input prompts} \\
        \toprule
        Prompt year & \multicolumn{6}{c}{Input prompt} & & \multicolumn{2}{c}{Correct output}\\
        \midrule
        2001 & \multicolumn{6}{l}{In 2001, the Enron [MASK] precipitated the passage of the Sarbanes-Oxley Act of 2002.} & & \multicolumn{2}{c}{scandal}  \\
        2003 & \multicolumn{6}{l}{In 2003, the [MASK] outbreak led to a major public health crisis.}  & & \multicolumn{2}{c}{SARS} \\
        2008 & \multicolumn{6}{l}{In 2008, the subprime mortgage [MASK] severely impacted the global economy.}  & & \multicolumn{2}{c}{crisis}  \\
        2016 & \multicolumn{6}{l}{In 2016, the uncertainty surrounding the [MASK] referendum led to heightened market volatility.}  & & \multicolumn{2}{c}{Brexit/EU} \\
        2020 & \multicolumn{6}{l}{In 2020, the [MASK] pandemic severely impacted the global economy.}  & & \multicolumn{2}{c}{coronavirus} \\
        2022 & \multicolumn{6}{l}{In 2022, the public release of [MASK]GPT marked a major milestone for generative AI.}  & & \multicolumn{2}{c}{Chat}  \\
        \bottomrule
        \\
        \multicolumn{10}{l}{Panel B: Output predictions} \\
        \toprule
        & \multicolumn{6}{c}{Prompt year} & & \multicolumn{2}{c}{Accuracy}\\
                \cmidrule{2-7} \cmidrule{9-10}
         & 2001 & 2003 & 2008 & 2016 & 2020 & 2022 & & Pre-cutoff & Post-cutoff \\
         \midrule
                 Correct output & scandal & SARS & crisis & Brexit/EU & coronavirus & Chat\\

        \midrule
        BERT              & \textcolor{blue}{\textbf{scandal}}  & flu  & \textcolor{blue}{\textbf{crisis}}  & \textcolor{blue}{\textbf{EU}}  &\cellcolor{lightgray}flu  &\cellcolor{lightgray}the  &  & $3/4$ & \cellcolor{lightgray}$0/2$\\
        ModernBERT              & \textcolor{blue}{\textbf{scandal}}  & \textcolor{blue}{\textbf{SARS}}  & \textcolor{blue}{\textbf{crisis}}  & \textcolor{blue}{\textbf{Brexit}}  & \textcolor{blue}{\textbf{coronavirus}}  & Auto  &  & $5/6$ & \cellcolor{lightgray}---\\
        \midrule
        $\text{ChronoBERT}_{1999}$   & \cellcolor{lightgray}War & \cellcolor{lightgray}influenza & \cellcolor{lightgray}tax & \cellcolor{lightgray}presidential & \cellcolor{lightgray}influenza & \cellcolor{lightgray}  &  & --- & \cellcolor{lightgray}$0/6$\\
        $\text{ChronoBERT}_{2000}$   & \cellcolor{lightgray}War & \cellcolor{lightgray}Ebola & \cellcolor{lightgray}market & \cellcolor{lightgray}presidential & \cellcolor{lightgray}influenza & \cellcolor{lightgray}O  &  & --- & \cellcolor{lightgray}$0/6$\\
        $\text{ChronoBERT}_{2001}$   & War & \cellcolor{lightgray}Ebola & \cellcolor{lightgray}market & \cellcolor{lightgray}presidential & \cellcolor{lightgray}influenza & \cellcolor{lightgray}p  &  & $0/1$ & \cellcolor{lightgray}$0/5$\\
        $\text{ChronoBERT}_{2002}$   & Amendment & \cellcolor{lightgray}Ebola & \cellcolor{lightgray}rate & \cellcolor{lightgray}election & \cellcolor{lightgray}influenza & \cellcolor{lightgray}EX  &  & $0/1$ & \cellcolor{lightgray}$0/5$\\
        $\text{ChronoBERT}_{2003}$   & \textcolor{blue}{\textbf{scandal}} & influenza & \cellcolor{lightgray}rate & \cellcolor{lightgray}Palestinian & \cellcolor{lightgray}influenza & \cellcolor{lightgray}E  &  & $1/2$ & \cellcolor{lightgray}$0/4$\\
        $\text{ChronoBERT}_{2004}$   & \textcolor{blue}{\textbf{scandal}} & \textcolor{blue}{\textbf{SARS}} & \cellcolor{lightgray}rate & \cellcolor{lightgray}tax & \cellcolor{lightgray}influenza & \cellcolor{lightgray}SU  &  & $2/2$ & \cellcolor{lightgray}$0/4$\\
        $\text{ChronoBERT}_{2005}$   & \textcolor{blue}{\textbf{scandal}} & \textcolor{blue}{\textbf{SARS}} & \cellcolor{lightgray}rate & \cellcolor{lightgray}presidential & \cellcolor{lightgray}influenza & \cellcolor{lightgray}O  &  & $2/2$ & \cellcolor{lightgray}$0/4$\\
        $\text{ChronoBERT}_{2006}$   & \textcolor{blue}{\textbf{scandal}} & \textcolor{blue}{\textbf{SARS}} & \cellcolor{lightgray}credit & \cellcolor{lightgray}presidential & \cellcolor{lightgray}influenza & \cellcolor{lightgray}O  &  & $2/2$ & \cellcolor{lightgray}$0/4$\\
        $\text{ChronoBERT}_{2007}$   & \textcolor{blue}{\textbf{scandal}} & \textcolor{blue}{\textbf{SARS}} & \cellcolor{lightgray}program & \cellcolor{lightgray}presidential & \cellcolor{lightgray}influenza & \cellcolor{lightgray}AR  &  & $2/2$ & \cellcolor{lightgray}$0/4$\\
        $\text{ChronoBERT}_{2008}$   & \textcolor{blue}{\textbf{scandal}} & \textcolor{blue}{\textbf{SARS}} & \textcolor{blue}{\textbf{crisis}} & \cellcolor{lightgray}Kyoto & \cellcolor{lightgray}influenza & \cellcolor{lightgray}AR  &  & $3/3$ & \cellcolor{lightgray}$0/3$\\
        $\text{ChronoBERT}_{2009}$   & \textcolor{blue}{\textbf{scandal}} & \textcolor{blue}{\textbf{SARS}} & \textcolor{blue}{\textbf{crisis}} & \cellcolor{lightgray}Kyoto & \cellcolor{lightgray}influenza & \cellcolor{lightgray}E  &  & $3/3$ & \cellcolor{lightgray}$0/3$\\
        $\text{ChronoBERT}_{2010}$   & \textcolor{blue}{\textbf{scandal}} & \textcolor{blue}{\textbf{SARS}} & \textcolor{blue}{\textbf{crisis}} & \cellcolor{lightgray}tax & \cellcolor{lightgray}influenza & \cellcolor{lightgray}AR  &  & $3/3$ & \cellcolor{lightgray}$0/3$\\
        $\text{ChronoBERT}_{2011}$   & \textcolor{blue}{\textbf{scandal}} & \textcolor{blue}{\textbf{SARS}} & \textcolor{blue}{\textbf{crisis}} & \cellcolor{lightgray}Iraq & \cellcolor{lightgray}influenza & \cellcolor{lightgray}U  &  & $3/3$ & \cellcolor{lightgray}$0/3$\\
        $\text{ChronoBERT}_{2012}$   & \textcolor{blue}{\textbf{scandal}} & \textcolor{blue}{\textbf{SARS}} & \textcolor{blue}{\textbf{crisis}} & \cellcolor{lightgray}presidential & \cellcolor{lightgray}influenza & \cellcolor{lightgray}AR  &  & $3/3$ & \cellcolor{lightgray}$0/3$\\
        $\text{ChronoBERT}_{2013}$   & \textcolor{blue}{\textbf{scandal}} & \textcolor{blue}{\textbf{SARS}} & \textcolor{blue}{\textbf{crisis}} & \cellcolor{lightgray}voter & \cellcolor{lightgray}influenza & \cellcolor{lightgray}e  &  & $3/3$ & \cellcolor{lightgray}$0/3$\\
        $\text{ChronoBERT}_{2014}$   & \textcolor{blue}{\textbf{scandal}} & \textcolor{blue}{\textbf{SARS}} & \textcolor{blue}{\textbf{crisis}} & \cellcolor{lightgray}Iraq & \cellcolor{lightgray}influenza & \cellcolor{lightgray}i  &  & $3/3$ & \cellcolor{lightgray}$0/3$\\
        $\text{ChronoBERT}_{2015}$   & \textcolor{blue}{\textbf{scandal}} & \textcolor{blue}{\textbf{SARS}} & \textcolor{blue}{\textbf{crisis}} & \cellcolor{lightgray}presidential & \cellcolor{lightgray}Ebola & \cellcolor{lightgray}i  &  & $3/3$ & \cellcolor{lightgray}$0/3$\\
        $\text{ChronoBERT}_{2016}$   & \textcolor{blue}{\textbf{scandal}} & \textcolor{blue}{\textbf{SARS}} & \textcolor{blue}{\textbf{crisis}} & \textcolor{blue}{\textbf{Brexit}} & \cellcolor{lightgray}influenza & \cellcolor{lightgray}i  & & $4/4$ & \cellcolor{lightgray}$0/2$\\
        $\text{ChronoBERT}_{2017}$   & \textcolor{blue}{\textbf{scandal}} & Ebola & \textcolor{blue}{\textbf{crisis}} & \textcolor{blue}{\textbf{Brexit}} & \cellcolor{lightgray}Ebola & \cellcolor{lightgray}X &  & $3/4$ & \cellcolor{lightgray}$0/2$\\
        $\text{ChronoBERT}_{2018}$   & \textcolor{blue}{\textbf{scandal}} & Ebola & \textcolor{blue}{\textbf{crisis}} & \textcolor{blue}{\textbf{Brexit}} & \cellcolor{lightgray}Ebola & \cellcolor{lightgray}X &  & $3/4$ & \cellcolor{lightgray}$0/2$\\
        $\text{ChronoBERT}_{2019}$   & \textcolor{blue}{\textbf{scandal}} & \textcolor{blue}{\textbf{SARS}} & \textcolor{blue}{\textbf{crisis}} & \textcolor{blue}{\textbf{Brexit}} & \cellcolor{lightgray}influenza & \cellcolor{lightgray}W &  & $4/4$ & \cellcolor{lightgray}$0/2$\\
        $\text{ChronoBERT}_{2020}$   & \textcolor{blue}{\textbf{scandal}} & \textcolor{blue}{\textbf{SARS}} & \textcolor{blue}{\textbf{crisis}} & \textcolor{blue}{\textbf{Brexit}} & \textcolor{blue}{\textbf{coronavirus}} & \cellcolor{lightgray}V &  & $5/5$ & \cellcolor{lightgray}$0/1$\\
        $\text{ChronoBERT}_{2021}$   & \textcolor{blue}{\textbf{scandal}} & Ebola & \textcolor{blue}{\textbf{crisis}} & \textcolor{blue}{\textbf{Brexit}} & \textcolor{blue}{\textbf{coronavirus}} & \cellcolor{lightgray}X  & & $4/5$ & \cellcolor{lightgray}$0/1$\\
        $\text{ChronoBERT}_{2022}$   & \textcolor{blue}{\textbf{scandal}} & Ebola & \textcolor{blue}{\textbf{crisis}} & \textcolor{blue}{\textbf{Brexit}} & \textcolor{blue}{\textbf{coronavirus}} & \#  & & $4/6$ & \cellcolor{lightgray}---\\
        $\text{ChronoBERT}_{2023}$   & \textcolor{blue}{\textbf{scandal}} & \textcolor{blue}{\textbf{SARS}} & \textcolor{blue}{\textbf{crisis}} & \textcolor{blue}{\textbf{Brexit}} & \textcolor{blue}{\textbf{coronavirus}} & \textcolor{blue}{\textbf{Chat}}  & & $6/6$ & \cellcolor{lightgray}---\\
        $\text{ChronoBERT}_{2024}$   & \textcolor{blue}{\textbf{scandal}} & \textcolor{blue}{\textbf{SARS}} & \textcolor{blue}{\textbf{crisis}} & \textcolor{blue}{\textbf{Brexit}} & \textcolor{blue}{\textbf{coronavirus}} & \textcolor{blue}{\textbf{Chat}}  & & $6/6$ & \cellcolor{lightgray}---\\
        \hline
        \multicolumn{7}{l}{$\text{ChronoBERT}_{1999}$ through $\text{ChronoBERT}_{2024}$}  & & $72/80$ & \cellcolor{lightgray}$0/76$\\
        \bottomrule
    \end{tabular}
    }
    \caption{Masked-Token Predictions of Major Events using ChronoBERT}
    \label{tab:majorevents_chronobert}
        \bigskip
    \footnotesize
This table reports ChronoBERT's masked-token predictions for sentences describing major events from various years.
Panel A shows the input prompts for each event. Panel B shows the output predictions for each event by each model in the ChronoBERT series.
For each masked token, the model's prediction is the token with the highest probability, chosen deterministically. The gray-shaded area denotes predictions covering the post-knowledge cutoff period. The blue text highlights correct predictions. For comparison, predictions from BERT (released in 2018) and ModernBERT (released in 2024) are also included.
\end{table}

\begin{table}[!htb]
    \resizebox{\textwidth}{!}{
    \begin{tabular}{lccccccccc}
        \multicolumn{10}{l}{Panel A: Input prompts} \\
        \toprule
        Prompt year & \multicolumn{6}{c}{Input prompt} & & \multicolumn{2}{c}{Correct output}\\
        \midrule
        2001 & \multicolumn{6}{l}{The Sarbanes-Oxley Act was introduced in response to the 2001 Enron \rule{2cm}{0.1mm}} & & \multicolumn{2}{c}{scandal}  \\
        2003 & \multicolumn{6}{l}{In 2003, a major public health crisis was the outbreak of the virus known as \rule{2cm}{0.1mm}}  & & \multicolumn{2}{c}{SARS} \\
        2008 & \multicolumn{6}{l}{In 2008, the global economy was dominated by the subprime mortgage \rule{2cm}{0.1mm}}  & & \multicolumn{2}{c}{crisis}  \\
        2016 & \multicolumn{6}{l}{In 2016, market volatility increased due to the uncertainty surrounding the vote in the UK, known as the \rule{2cm}{0.1mm}}  & & \multicolumn{2}{c}{Brexit referendum/vote} \\
        2020 & \multicolumn{6}{l}{In 2020, the global economy was severely impacted by the pandemic known as \rule{2cm}{0.1mm}}  & & \multicolumn{2}{c}{COVID} \\
        2022 & \multicolumn{6}{l}{In 2022, a major milestone for generative AI was marked by the public release of the AI assistant known as \rule{2cm}{0.1mm}}  & & \multicolumn{2}{c}{ChatGPT}  \\
        \bottomrule
        \\
        \multicolumn{10}{l}{Panel B: Output predictions} \\
        \toprule
        & \multicolumn{6}{c}{Prompt year} & & \multicolumn{2}{c}{Accuracy}\\
                \cmidrule{2-7} \cmidrule{9-10}
         & 2001 & 2003 & 2008 & 2016 & 2020 & 2022 & & Pre-cutoff & Post-cutoff \\
        \midrule
        Correct output & scandal & SARS & crisis & Brexit referendum/vote & COVID & ChatGPT\\
        \midrule
        GPT-2              & \textcolor{blue}{\textbf{scandal}.\textbackslash n}  & chikung  & \textcolor{blue}{\textbf{crisis}, which}  & \textcolor{blue}{\textbf{Brexit vote}.}  &\cellcolor{lightgray}the Ebola virus  &\cellcolor{lightgray} the ``AI &  & $3/4$ & \cellcolor{lightgray}$0/2$\\
        GPT-2 XL              & \textcolor{blue}{\textbf{scandal}, in}  & \textcolor{blue}{\textbf{SARS},}   & \textcolor{blue}{\textbf{crisis}. The}  & \textcolor{blue}{\textbf{Brexit}, and}  &\cellcolor{lightgray}the ``Great  &\cellcolor{lightgray} Alexa. Alexa &  & $4/4$ & \cellcolor{lightgray}$0/2$\\
        \midrule
        $\text{ChronoGPT}_{1999}$   & \cellcolor{lightgray}ous Claims Act & \cellcolor{lightgray}the Ebola virus & \cellcolor{lightgray}market. The & \cellcolor{lightgray}``Budget & \cellcolor{lightgray} the ``green & \cellcolor{lightgray} the AI assistant  &  & --- & \cellcolor{lightgray}$0/6$\\
        $\text{ChronoGPT}_{2000}$   & \cellcolor{lightgray}a-H & \cellcolor{lightgray}Ebola virus in & \cellcolor{lightgray}rate, which & \cellcolor{lightgray}\textquotedblright B & \cellcolor{lightgray} the Ebola virus & \cellcolor{lightgray} AI-R  &  & --- & \cellcolor{lightgray}$0/6$\\
        $\text{ChronoGPT}_{2001}$   & Agreement, which & \cellcolor{lightgray}Ebola (E & \cellcolor{lightgray}market & \cellcolor{lightgray}\textquotedblright V & \cellcolor{lightgray}the Ebola virus & \cellcolor{lightgray}  AI Magazine.  &  & $0/1$ & \cellcolor{lightgray}$0/5$\\
        $\text{ChronoGPT}_{2002}$   & \textcolor{blue}{\textbf{scandal}. The} & \cellcolor{lightgray}hantav & \cellcolor{lightgray}market, which & \cellcolor{lightgray}\textquotedblright V & \cellcolor{lightgray} the ``Asian & \cellcolor{lightgray} the AI-R  &  & $1/1$ & \cellcolor{lightgray}$0/5$\\
        $\text{ChronoGPT}_{2003}$   & \textcolor{blue}{\textbf{scandal}. It} & West Nile virus& \cellcolor{lightgray}market, which & \cellcolor{lightgray}``Voting & \cellcolor{lightgray} the Ebola virus & \cellcolor{lightgray} AIBO.  &  & $1/2$ & \cellcolor{lightgray}$0/4$\\
        $\text{ChronoGPT}_{2004}$   & \textcolor{blue}{\textbf{scandal}. It} & \textcolor{blue}{\textbf{SARS}.} & \cellcolor{lightgray}market, which & \cellcolor{lightgray} ``Voting & \cellcolor{lightgray} the SARS & \cellcolor{lightgray} AIBO.  &  & $2/2$ & \cellcolor{lightgray}$0/4$\\
        $\text{ChronoGPT}_{2005}$   & \textcolor{blue}{\textbf{scandal}. It} & \textcolor{blue}{\textbf{SARS} (} & \cellcolor{lightgray}market, which & \cellcolor{lightgray} ``London effect & \cellcolor{lightgray} SARS. & \cellcolor{lightgray} AIBO.  &  & $2/2$ & \cellcolor{lightgray}$0/4$\\
        $\text{ChronoGPT}_{2006}$   & \textcolor{blue}{\textbf{scandal}. It} & \textcolor{blue}{\textbf{SARS},} & \cellcolor{lightgray}market, which & \cellcolor{lightgray} ``London effect & \cellcolor{lightgray} the SARS & \cellcolor{lightgray} the AIBO  &  & $2/2$ & \cellcolor{lightgray}$0/4$\\
        $\text{ChronoGPT}_{2007}$   & \textcolor{blue}{\textbf{scandal}. It} & \textcolor{blue}{\textbf{SARS},} & \cellcolor{lightgray}market, which & \cellcolor{lightgray} ``Budget & \cellcolor{lightgray} the Black Death & \cellcolor{lightgray} the AIBO   &  & $2/2$ & \cellcolor{lightgray}$0/4$\\
        $\text{ChronoGPT}_{2008}$   & \textcolor{blue}{\textbf{scandal}. It} & \textcolor{blue}{\textbf{SARS},} & \textcolor{blue}{\textbf{crisis}, which} & \cellcolor{lightgray} ``Voting & \cellcolor{lightgray} the Black Death & \cellcolor{lightgray} the AIBO   &  & $3/3$ & \cellcolor{lightgray}$0/3$\\
        $\text{ChronoGPT}_{2009}$   & \textcolor{blue}{\textbf{scandal}. The} & \textcolor{blue}{\textbf{SARS},} & \textcolor{blue}{\textbf{crisis}, which} & \cellcolor{lightgray} ``UK vote & \cellcolor{lightgray} the Black Death & \cellcolor{lightgray} the ``C  &  & $3/3$ & \cellcolor{lightgray}$0/3$\\
        $\text{ChronoGPT}_{2010}$   & \textcolor{blue}{\textbf{scandal}. The} & \textcolor{blue}{\textbf{SARS},} & \textcolor{blue}{\textbf{crisis}, which} & \cellcolor{lightgray} ``UK vote & \cellcolor{lightgray} the SARS & \cellcolor{lightgray} the AIBO  &  & $3/3$ & \cellcolor{lightgray}$0/3$\\
        $\text{ChronoGPT}_{2011}$   & \textcolor{blue}{\textbf{scandal}. The} & \textcolor{blue}{\textbf{SARS},} & \textcolor{blue}{\textbf{crisis}, which} & \cellcolor{lightgray} ``Budget & \cellcolor{lightgray} the H1 & \cellcolor{lightgray} the AIBO   &  & $3/3$ & \cellcolor{lightgray}$0/3$\\
        $\text{ChronoGPT}_{2012}$   & \textcolor{blue}{\textbf{scandal}. The} & \textcolor{blue}{\textbf{SARS},} & \textcolor{blue}{\textbf{crisis}, which} & \cellcolor{lightgray} ``Budget & \cellcolor{lightgray} the Black Death & \cellcolor{lightgray} the AI-  &  & $3/3$ & \cellcolor{lightgray}$0/3$\\
        $\text{ChronoGPT}_{2013}$   & \textcolor{blue}{\textbf{scandal}. The} & \textcolor{blue}{\textbf{SARS},} & \textcolor{blue}{\textbf{crisis}, which} & \cellcolor{lightgray} ``Budget & \cellcolor{lightgray} the "Asian & \cellcolor{lightgray} the AIBO   &  & $3/3$ & \cellcolor{lightgray}$0/3$\\
        $\text{ChronoGPT}_{2014}$   & \textcolor{blue}{\textbf{scandal}. The} & \textcolor{blue}{\textbf{SARS},} & \textcolor{blue}{\textbf{crisis}, which} & \cellcolor{lightgray} ``Voting & \cellcolor{lightgray} the "Asian & \cellcolor{lightgray} the ``C  &  & $3/3$ & \cellcolor{lightgray}$0/3$\\
        $\text{ChronoGPT}_{2015}$   & \textcolor{blue}{\textbf{scandal}. The} & \textcolor{blue}{\textbf{SARS},} & \textcolor{blue}{\textbf{crisis}, which} & \cellcolor{lightgray} ``UK vote & \cellcolor{lightgray} the Ebola virus & \cellcolor{lightgray} the AIBO  &  & $3/3$ & \cellcolor{lightgray}$0/3$\\
        $\text{ChronoGPT}_{2016}$   & \textcolor{blue}{\textbf{scandal}. The} & \textcolor{blue}{\textbf{SARS},} & \textcolor{blue}{\textbf{crisis}, which} & ``UK vote & \cellcolor{lightgray} the Ebola virus & \cellcolor{lightgray} the ``C  & & $3/4$ & \cellcolor{lightgray}$0/2$\\
        $\text{ChronoGPT}_{2017}$   & \textcolor{blue}{\textbf{scandal}. The} & \textcolor{blue}{\textbf{SARS},} & \textcolor{blue}{\textbf{crisis}, which} & \textcolor{blue}{\textbf{Brexit referendum}.} & \cellcolor{lightgray} the Ebola virus & \cellcolor{lightgray} the ``C &  & $4/4$ & \cellcolor{lightgray}$0/2$\\
        $\text{ChronoGPT}_{2018}$   & \textcolor{blue}{\textbf{scandal}. The} & Ebola in West & \textcolor{blue}{\textbf{crisis}, which} & \textcolor{blue}{\textbf{Brexit vote}.} & \cellcolor{lightgray} the Ebola virus & \cellcolor{lightgray} DeepMind's &  & $3/4$ & \cellcolor{lightgray}$0/2$\\
        $\text{ChronoGPT}_{2019}$   & \textcolor{blue}{\textbf{scandal}. The} & Ebola in West & \textcolor{blue}{\textbf{crisis}, which} & \textcolor{blue}{\textbf{Brexit referendum}.} & \cellcolor{lightgray} the Ebola virus & \cellcolor{lightgray} Siri. The &  & $3/4$ & \cellcolor{lightgray}$0/2$\\
        $\text{ChronoGPT}_{2020}$   & \textcolor{blue}{\textbf{scandal}. The} & \textcolor{blue}{\textbf{SARS}-} & \textcolor{blue}{\textbf{crisis}, which} & \textcolor{blue}{\textbf{Brexit referendum}.} & \textcolor{blue}{\textbf{COVID-}} & \cellcolor{lightgray} Siri. The &  & $5/5$ & \cellcolor{lightgray}$0/1$\\
        $\text{ChronoGPT}_{2021}$   & \textcolor{blue}{\textbf{scandal}. The} & \textcolor{blue}{\textbf{SARS}-} & \textcolor{blue}{\textbf{crisis}, which} & \textcolor{blue}{\textbf{Brexit referendum}.} & \textcolor{blue}{\textbf{COVID-}} & \cellcolor{lightgray} Alexa. The  & & $5/5$ & \cellcolor{lightgray}$0/1$\\
        $\text{ChronoGPT}_{2022}$   & \textcolor{blue}{\textbf{scandal}. The} & \textcolor{blue}{\textbf{SARS}-} & \textcolor{blue}{\textbf{crisis}, which} & \textcolor{blue}{\textbf{Brexit referendum}.} & \textcolor{blue}{\textbf{COVID-}} &  Alexa. The  & & $5/6$ & \cellcolor{lightgray}---\\
        $\text{ChronoGPT}_{2023}$   & \textcolor{blue}{\textbf{scandal}. The} & \textcolor{blue}{\textbf{SARS}-} & \textcolor{blue}{\textbf{crisis}, which} & \textcolor{blue}{\textbf{Brexit referendum}.} & \textcolor{blue}{\textbf{COVID-}} & \textcolor{blue}{\textbf{ChatGPT}}  & & $6/6$ & \cellcolor{lightgray}---\\
        $\text{ChronoGPT}_{2024}$   & \textcolor{blue}{\textbf{scandal}. The} & \textcolor{blue}{\textbf{SARS}-} & \textcolor{blue}{\textbf{crisis}, which} & \textcolor{blue}{\textbf{Brexit referendum}.} & \textcolor{blue}{\textbf{COVID-}} & \textcolor{blue}{\textbf{ChatGPT}}  & & $6/6$ & \cellcolor{lightgray}---\\
        \hline
        \multicolumn{7}{l}{$\text{ChronoGPT}_{1999}$ through $\text{ChronoGPT}_{2024}$}  & & $74/80$ & \cellcolor{lightgray}$0/76$\\
        \bottomrule
    \end{tabular}
    }
    \caption{Next-Token Predictions of Major Events using ChronoGPT}
    \label{tab:majorevents_chronogpt}
        \bigskip
    \footnotesize
This table reports ChronoGPT's next-token predictions for sentences describing major events from various years.
Panel A shows the input prompts for each event. Panel B shows the output predictions for each event by each model in the ChronoGPT series.
Each sequence includes exactly three predicted tokens, with the most probable token selected deterministically at each step.
The gray-shaded area denotes predictions covering the post-knowledge cutoff period. The blue text highlights correct predictions. For comparison, predictions from GPT-2 and GPT-2 XL (released in 2019) are also included.
\end{table}

In addition to identifying the names of presidents, consider a test that involves identifying significant events across different years. For each event, we construct a sentence with a missing token that represents a defining element of the event and is the target of the prediction. The sentences cover the Enron \emph{scandal}, the \emph{SARS} outbreak, the subprime mortgage \emph{crisis}, the \emph{Brexit} referendum, the \emph{coronavirus}/\emph{COVID} pandemic, and the public release of \emph{ChatGPT}. 
% The missing, event-defining tokens---the subjects of our prediction exercise---are presented in quotation marks. 

Table \ref{tab:majorevents_chronobert} presents the masked-token predictions using ChronoBERT, and Table \ref{tab:majorevents_chronogpt} presents the next-token predictions using ChronoGPT. The specific textual sequences containing the event-defining missing tokens are reported in each table. In the pre-cutoff period represented by the non-shaded area, ChronoBERT correctly makes a majority of predictions (72 out of 80), outperforming BERT (3 out of 4) and ModernBERT (5 out of 6). ChronoGPT also correctly makes most predictions (74 out of 80), outperforming GPT-2 (3 out of 4), though slightly below GPT-2 XL's perfect performance. None of the ChronoBERT and ChronoGPT models correctly predicts an event-defining token from a future event, as represented by the gray-shaded area. Overall, these findings validate that the textual data used to train our chronologically consistent models contains no evidence of leakage.

\subsection{Predicting Stock Returns using Financial News}

\begin{table}[!htb]
\scriptsize
\begin{center}
\begin{tabularx}{\textwidth}{@{\hskip\tabcolsep\extracolsep\fill}l*{12}r}
\toprule
& \multicolumn{3}{c}{$\text{ChronoBERT}_{\text{Realtime}}$} & \multicolumn{3}{c}{$\text{ChronoGPT}_{\text{Realtime}}$} & \multicolumn{3}{c}{Llama 3.1} & \multicolumn{3}{c}{GPT-2 XL} \\
\cmidrule(lr){2-4} \cmidrule(lr){5-7} \cmidrule(lr){8-10} \cmidrule(lr){11-13}
& Mean & SD & SR & Mean & SD & SR & Mean & SD & SR & Mean & SD & SR \\
\midrule
Low (L)  & -23.30 & 25.86 & -0.90 & -19.82 & 25.15 & -0.79 & -23.71 & 26.15 & -0.91 & -22.25 & 25.75 & -0.86\\
2        & -2.43  & 25.20 & -0.10 &   1.48  & 25.48 &  0.06 & -4.77  & 25.31 & -0.19 & -1.36 & 25.32 & -0.05\\
3        &  4.17  & 25.64 &  0.16 &   0.61  & 24.87 &  0.02 & -0.24  & 24.86 & -0.01 & -1.10 & 24.82 & -0.04\\
4        &  4.17  & 24.58 &  0.17 &   2.41  & 24.26 &  0.10 &  3.84  & 24.62 &  0.16 & 2.60 & 24.43 & 0.11\\
5        &  3.94  & 24.22 &  0.16 &   7.18  & 24.54 &  0.29 &  7.47  & 24.65 &  0.30 & 6.24 & 24.30 & 0.26\\
6        & 10.81  & 24.13 &  0.45 &   5.78  & 24.20 &  0.24 & 12.03  & 24.23 &  0.50 & 10.14 & 24.53 & 0.41\\
7        & 14.56  & 24.23 &  0.60 &  10.82  & 23.84 &  0.45 & 13.31  & 24.33 &  0.55 & 16.90 & 24.76 & 0.68\\
8        & 16.38  & 23.64 &  0.69 &  15.74  & 24.30 &  0.65 & 15.13  & 23.79 &  0.64 & 15.69 & 24.09 & 0.65\\
9        & 23.95  & 24.45 &  0.98 &  22.63  & 24.19 &  0.94 & 24.68  & 23.88 &  1.03 & 24.85 & 24.24 & 1.03\\
High (H) & 37.71  & 24.53 &  1.54 &  43.08  & 25.41 &  1.70 & 42.20  & 25.05 &  1.68 & 38.25 & 24.53 & 1.56\\
H-L      & 61.02  & 12.72 &  4.80 &  62.90  & 12.78 &  4.92 & 65.91  & 13.46 &  4.90 & 60.51 & 13.24 & 4.57\\\midrule
& \multicolumn{3}{c}{ModernBERT} & \multicolumn{3}{c}{BERT} & \multicolumn{3}{c}{FinBERT} & \multicolumn{3}{c}{StoriesLM} \\
\cmidrule(lr){2-4} \cmidrule(lr){5-7} \cmidrule(lr){8-10} \cmidrule(lr){11-13}
& Mean & SD & SR & Mean & SD & SR & Mean & SD & SR & Mean & SD & SR \\
\midrule
Low (L)  & -24.39 & 26.07 & -0.94 & -22.52 & 26.21 & -0.86 & -23.96 & 26.86 & -0.89 & -17.80 & 26.52 & -0.67 \\
2        &  0.49  & 25.86 &  0.02 & -5.05  & 25.55 & -0.20 & -3.17  & 25.64 & -0.12 & -1.19  & 25.26 & -0.05 \\
3        &  1.64  & 25.08 &  0.07 &  3.12  & 24.92 &  0.13 &  3.36  & 24.83 &  0.14 &  1.86  & 24.92 &  0.07 \\
4        &  5.41  & 24.81 &  0.22 &  8.14  & 24.62 &  0.33 &  7.19  & 24.52 &  0.29 &  5.90  & 24.62 &  0.24 \\
5        &  8.67  & 24.67 &  0.35 & 10.81  & 24.44 &  0.44 &  9.17  & 24.39 &  0.38 &  4.99  & 24.30 &  0.21 \\
6        & 10.02  & 23.99 &  0.42 &  9.38  & 24.02 &  0.39 & 11.47  & 24.03 &  0.48 & 11.88  & 23.90 &  0.50 \\
7        & 12.99  & 23.93 &  0.54 & 14.54  & 23.83 &  0.61 & 16.54  & 23.92 &  0.69 & 12.41  & 23.66 &  0.52 \\
8        & 16.17  & 23.69 &  0.68 & 18.51  & 24.04 &  0.77 & 19.16  & 23.65 &  0.81 & 18.93  & 24.19 &  0.78 \\
9        & 23.45  & 23.97 &  0.98 & 19.68  & 23.90 &  0.82 & 20.70  & 23.88 &  0.87 & 23.25  & 24.30 &  0.96 \\
High (H) & 35.53  & 24.42 &  1.45 & 33.37  & 24.88 &  1.34 & 29.51  & 24.60 &  1.20 & 29.73  & 24.78 &  1.20 \\
H-L      & 59.92  & 13.03 &  4.60 & 55.89  & 13.38 &  4.18 & 53.47  & 13.85 &  3.86 & 47.53  & 13.90 &  3.42 \\
\bottomrule
\end{tabularx}
\end{center}
\caption{Performance of the LLM Portfolios}
\label{tab:decile}
\bigskip
\small
This table presents annualized performance metrics (mean return, standard deviation, and Sharpe ratio) for decile portfolios sorted by next-day return predictions from financial news. Portfolios are rebalanced daily, with the "H-L" row representing a strategy of longing the top decile and shorting the bottom decile. All values are in percentage points except Sharpe ratios. All portfolios are equal-weighted. Data spans January 2008–July 2023.
\end{table}

To quantify the economic gains from enhanced language understanding, we analyze stock return predictions using news embeddings from different language models. We construct portfolios by sorting stocks based on each model's return forecasts and evaluate the performance of long-short spreads generated by these rankings.\footnote{Following \citet{he2024empirical}, we conduct a robustness check by forecasting the probability of a positive stock return on the subsequent trading day. The outcomes closely mirror those obtained from the return forecasts.}

Table \ref{tab:decile} presents the decile portfolio performance for realtime models ($\text{ChronoBERT}_{\text{Realtime}}$ and$\text{ChronoGPT}_{\text{Realtime}}$). For example, in the year $t$, we would use the latest available model checkpoint at year $t-1$ to embed the news articles in that year and make predictions based on the embeddings. For comparison, we also report six other benchmarks: (1) Llama-3.1-8B; (2) GPT-2 XL; (3) ModernBERT; (4) BERT; (5) FinBERT; and (6) StoriesLM. In this news return prediction setting, the H-L portfolios from $\text{ChronoBERT}_{\text{Realtime}}$ and $\text{ChronoGPT}_{\text{Realtime}}$ generate Sharpe ratios of 4.80 and 4.92, outperforming GPT-2 XL, ModernBERT, BERT, FinBERT and StoriesLM. These results demonstrate that increased language understanding indeed translates into significant economic gains.

\begin{table}[!htb]
    \scriptsize
    \setlength{\tabcolsep}{3pt}
    \begin{center}
    \begin{tabularx}{\textwidth}{@{\hskip\tabcolsep\extracolsep\fill}l*{9}c}
    \toprule
      & ChronoBERT & ChronoGPT & Llama 3.1 & GPT-2 XL & ModernBERT & BERT & FinBERT & StoriesLM \\
    \midrule
    ChronoBERT & 
    % & 0.076
    & 0.717 & 0.685 & 0.141 & 0.144 & 0.005 & 0.002 & 0.000 \\
    % ChronoGPT & 0.924 &  & 0.982 & 0.973 & 0.658 & 0.706 & 0.078 & 0.017 & 0.001 \\
    ChronoGPT & 0.283 
    %& 0.018 
    &  & 0.453 & 0.051 & 0.075 & 0.001 & 0.001 & 0.000 \\
    Llama 3.1 & 0.315 
    %& 0.027
    & 0.547 & & 0.048 & 0.069 & 0.001 & 0.000 & 0.000 \\
    GPT-2 XL & 0.859 
    %& 0.342 
    & 0.950 & 0.952 & & 0.555 & 0.038 & 0.010 & 0.000\\
    ModernBERT & 0.856 
    %& 0.294 
    & 0.925 & 0.931 & 0.445 & & 0.029 & 0.007 & 0.000 \\
    BERT & 0.995 
    %& 0.922
    & 0.999 & 0.999 &  0.962 & 0.971 &  & 0.116 & 0.005 \\
    FinBERT & 0.998 
    %& 0.983
    & 0.999 & 1.000 & 0.991 & 0.993 & 0.884 &  & 0.098 \\
    StoriesLM & 1.000 
    %& 0.999
    & 1.000 & 1.000 & 1.000 & 1.000 & 0.995 & 0.902 & \\
    \bottomrule
    \end{tabularx}
    \end{center}
    \caption{P-value of Pairwise Sharpe Ratio Difference Tests}\label{tab:sr_dff}
    \bigskip
    \small
    This table reports the p-value from the \citet{ledoit2008robust} Sharpe ratio difference test of the `H-L' portfolios from different LLMs in Table \ref{tab:decile}. Each entry corresponds to a test of the null hypothesis that the Sharpe ratio of the model in the row is smaller than or equal to that of the model in the column. The portfolio sample spans from January 2008 to July 2023.
\end{table}

We also report the p-value of the pairwise Sharpe ratio difference test using the \citet{ledoit2008robust} approach in Table \ref{tab:sr_dff}. Comparing realtime ChronoBERT and ChronoGPT against BERT, FinBERT and StoriesLM, we find that the models' improved language understanding and superior knowledge increases the investment Sharpe ratio, a difference that is both economically meaningful and statistically significant at the 1\% level. 

Comparing the investment performance of ChronoBERT and ChronoGPT against the Llama 3.1 model, we find that ChronoBERT and ChronoGPT generate comparable Sharpe ratios, with no statistically significant differences (p-value of 0.315 and 0.547). Between the two chronological models themselves, we observe no statistical differences (p-value of 0.283), indicating that both approaches effectively leverage temporal information.

These findings suggest that the performance of our chronologically consistent chrono models are comparable to powerful, larger-scale LLMs in this financial application, despite their limited training using historical textual data.

A somewhat surprising finding is that lookahead bias appears to be modest in this stock return forecasting application. The fact that the Llama-based performance is no different from that of our chronologically consistent models suggests understanding the news flow into the market in real-time is sufficient for generating substantial short-term returns, and that no disruption of the time continuum is required.

\begin{figure}[!htb]
    \begin{center}
    \includegraphics[width=0.75\linewidth]{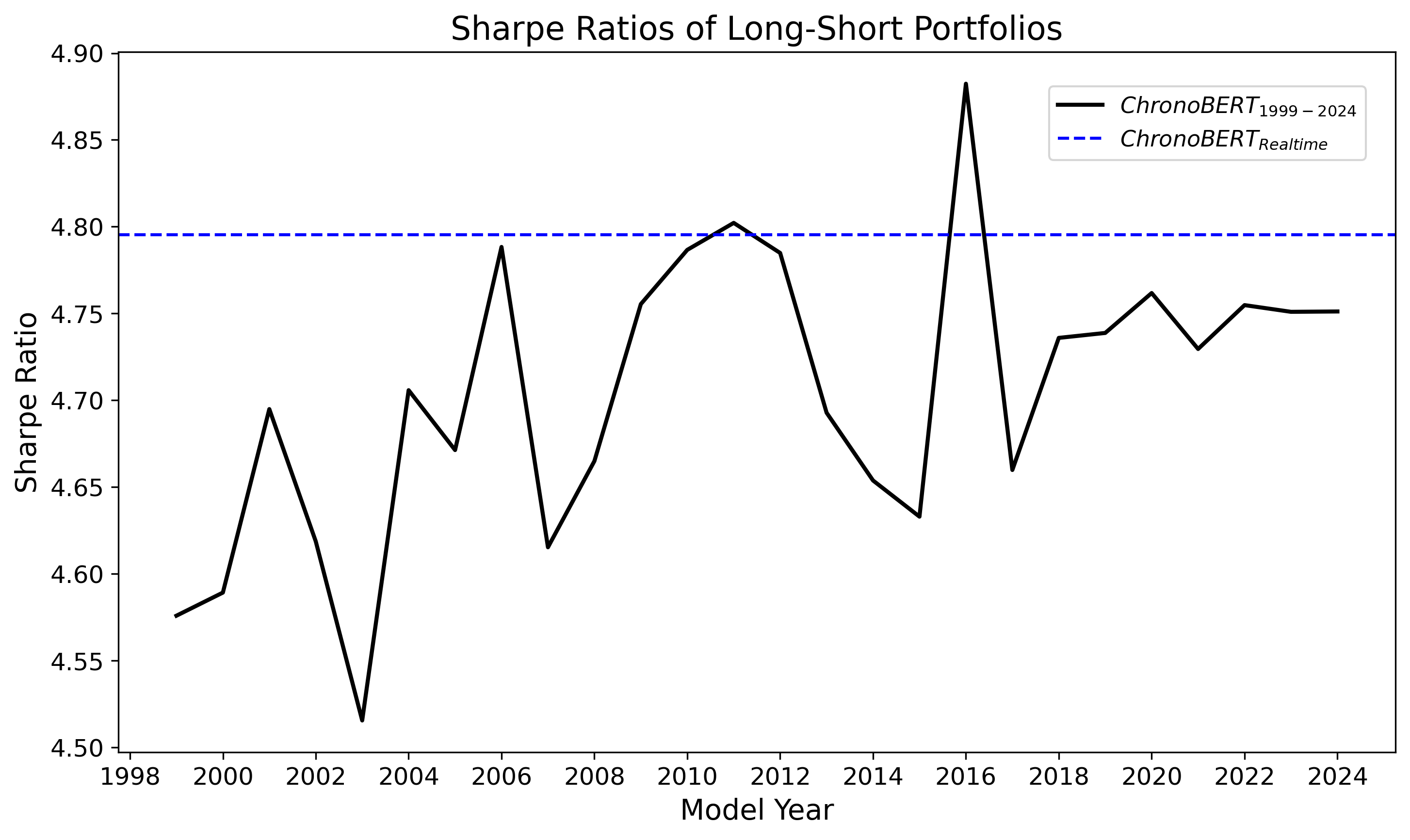}
    
    (a) ChronoBERT
     
    \par
    \bigskip
    
    \includegraphics[width=0.75\linewidth]{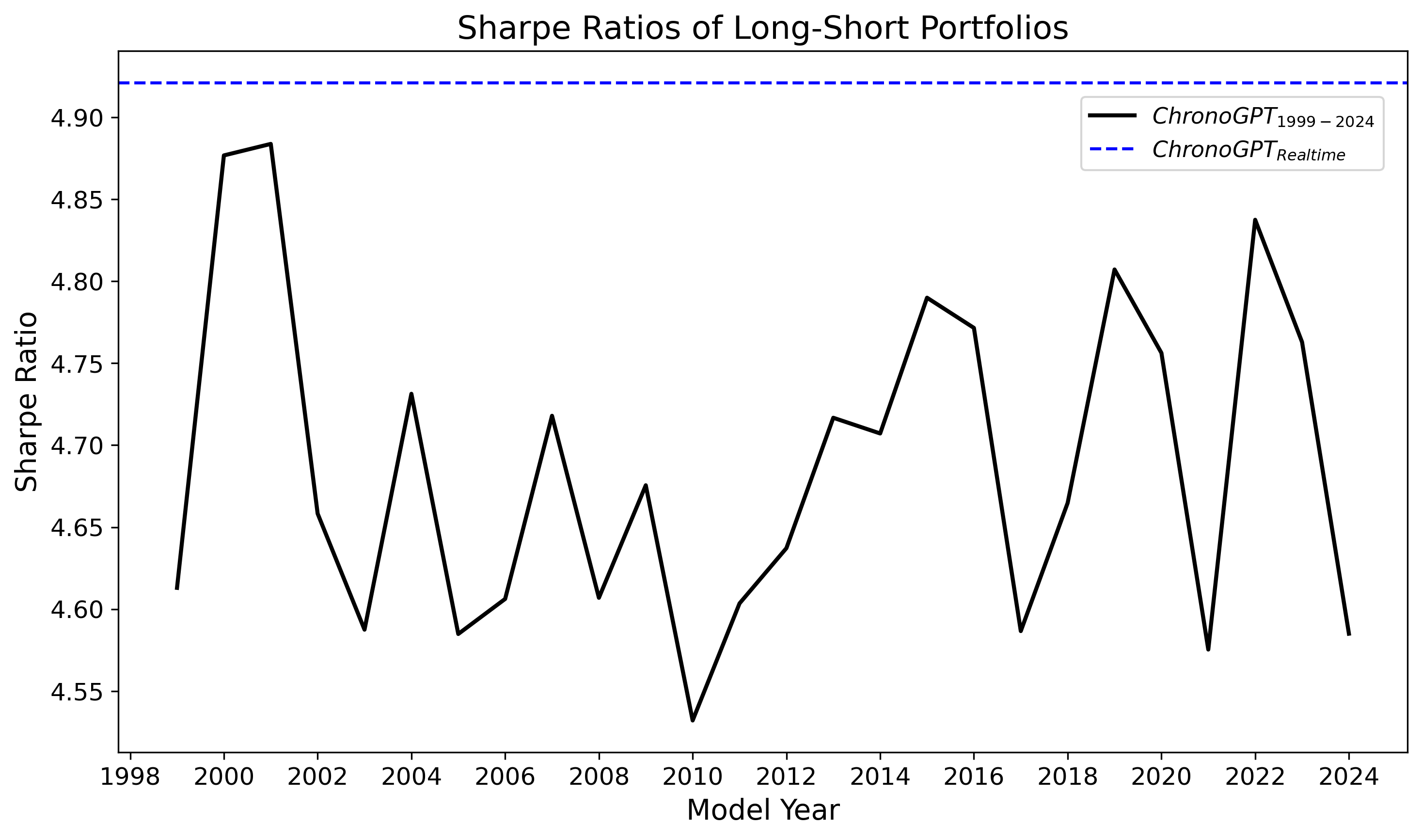}
    
    (b) ChronoGPT
    
    \end{center}
    \caption{Portfolios Performance across ChronoBERT and ChronoGPT Vintages}
    \label{fig:SR_time}
    \bigskip
    \small
    This figure illustrates the Sharpe ratios of long-short portfolios constructed using predictions derived from financial news, with language models pretrained on text data up to the time points indicated on the x-axis. The blue dashed line represents the performance of the chronologically consistent realtime models.  
    % The red line represents the Sharpe ratio of the Llama 3.1 model, while the blue line depicts the performance of the real-time ChronoBERT model.
\end{figure}

% \begin{figure}[!htb]
%     \begin{center}
%     \includegraphics[width=0.75\linewidth]{Figures/SR_time_gpt.png}\\
%     \end{center}
%     \caption{Portfolios Performance across ChronoGPT Vintages}
%     \label{fig:SR_time_gpt}
%     \bigskip
%     \small
%     This figure illustrates the Sharpe ratios of long-short portfolios constructed using predictions derived from financial news, with language models pretrained on text data up to the time points indicated on the x-axis. The blue dashed line represents the performance of the $\text{ChronoGPT}_{\text{Realtime}}$ model.
% \end{figure}

While our models demonstrate improved language understanding as they are trained on more data over time (Figure \ref{fig:language_time}), a crucial question is whether this translates into economic gains. To test this, we evaluate the trading performance of the entire series of chronologically consistent models.

If enhanced language ability is the bottleneck of performance in return prediction tasks, we would expect to see a monotonic increase in the Sharpe ratio as the models' knowledge cutoff dates advance. However, Figure \ref{fig:SR_time} reveals a surprisingly different outcome. For each time $t$, we use the corresponding model from that vintage to generate predictions from news embeddings. Instead of a steady increase, we observe a distinct ``envelope'' phenomenon across both ChronoBERT and ChronoGPT vintages. The real-time model's Sharpe ratio consistently outperforms most vintage models, even those with more up-to-date knowledge cutoff.

This result highlights two critical points. First, lookahead bias is minimal. If significant lookahead bias is present, the final 2024 model, with the most comprehensive knowledge, would have been the top performer across all periods. Second, further improvements in generic language ability add only marginal value for this task: the earliest model is already close to the performance frontier.

What drives the envelope? One plausible explanation is temporal alignment. A real-time model is calibrated to the statistical regularities, vocabulary, and market narratives that prevail in its own period, allowing it to weigh news appropriately. Later models view earlier articles through a future-biased lens. For example, expressions such as ``meme stocks'', ``supply-chain disruptions'', or ``Fed pivot'' carry period-specific meanings. Interpreting 2020 news with the semantics learned in 2024 misaligns those signals and erodes predictive accuracy.

% Figure \ref{fig:SR_time} further presents the trading performance of the whole series of chronologically consistent models. Specifically, for each time $t$ in the figure, we use the time $t$ model to embed news articles and run predictions using embeddings from the model. We find consistent performance across all models in the series. The results again highlight (1) the return prediction exercise has modest lookahead bias; (2) the enhanced language understanding shown in Figure \ref{fig:language_time} indeed translates into significant economic gains.

\begin{figure}[!htb]
    \begin{center}
    \includegraphics[width=0.75\linewidth]{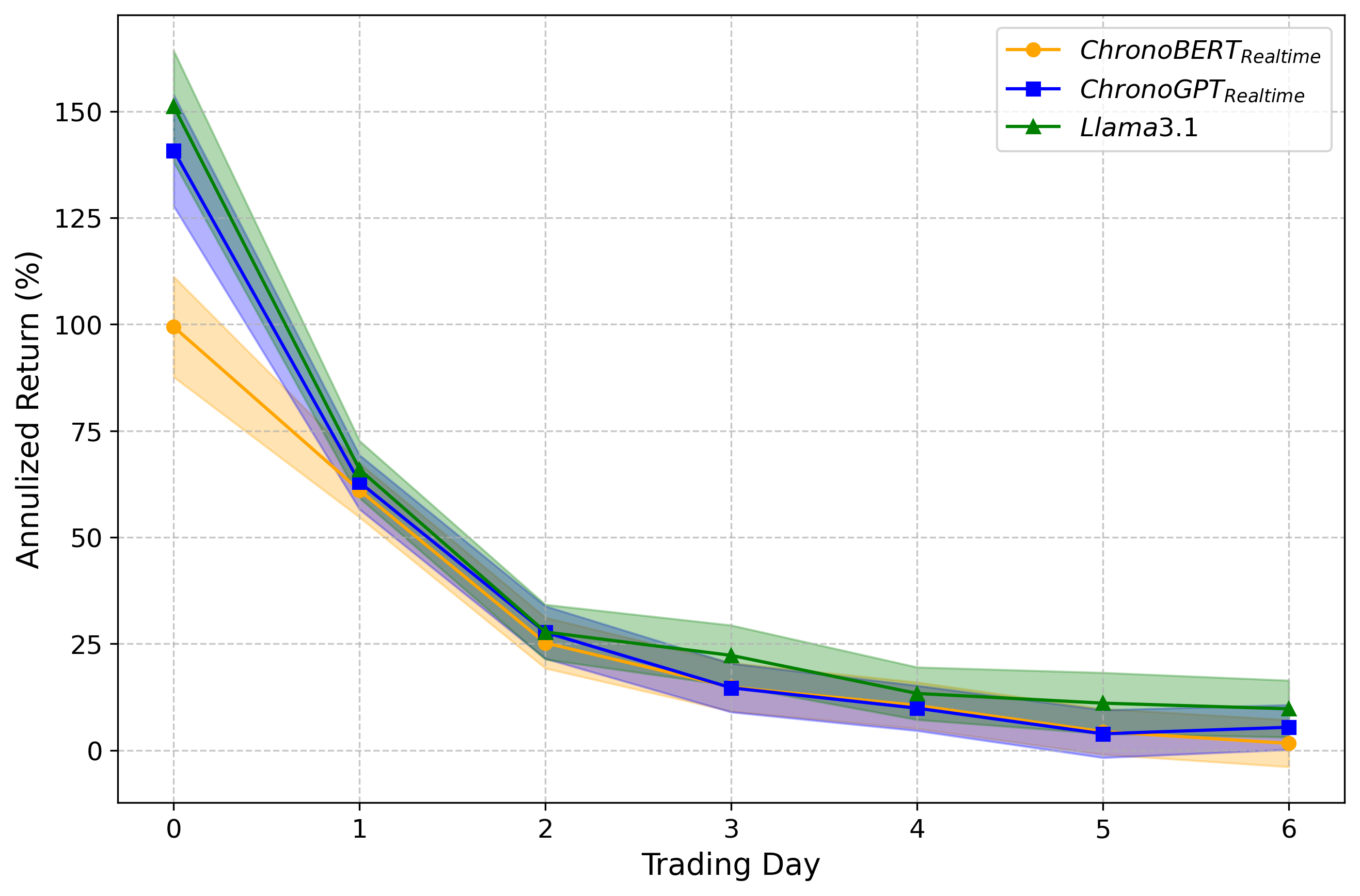}\\
    \end{center}
    \caption{Speed of News Incorporation}
    \label{fig:news_decay}
    \bigskip
    \small
    This figure compares the annualized returns (\%) of ChronoBERT, ChronoGPT and Llama 3.1 portfolios over a 7-day period. Day 0 marks the first trading day following the news announcement. The performance shown for these LLM portfolios in the analyses above specifically reflects their return on Day 1. Each point reflects the equal-weighted average return of a long-short decile portfolio formed by sorting stocks based on the respective language model’s next-day return predictions. Shaded areas represent 95\% confidence intervals.
\end{figure}

Figure \ref{fig:news_decay} complements our return-based analysis by illustrating the speed of news decay across models. We track the returns of long-short portfolios from Day 0 (the first trading day after a news release) through Day 6. All three models exhibit a steep decline in return predictability over the first two days, suggesting that most of the value from news-based predictions is rapidly incorporated into prices. $\text{ChronoGPT}_{\text{Realtime}}$ and Llama 3.1 generate the most concentrated Day 0 returns, while $\text{ChronoBERT}_{\text{Realtime}}$ exhibits a slightly more gradual decay pattern. The convergence of return profiles across all models by Days 3-6 indicates market efficiency in incorporating news information, regardless of model architecture. The decay pattern is also suggestive of market underreaction to news signals. Taken together, the findings suggest that while more sophisticated language understanding can amplify initial signal strength, the economic value of news-derived predictions exhibits a universal decay function.

\section{Conclusion}\label{sec:conclusion}

In this paper, we address the critical challenge of lookahead and training leakage in LLMs used in social science applications, particularly in financial forecasting. By introducing ChronoBERT and ChronoGPT, two series of chronologically consistent language models trained on timestamped text data, we demonstrate that chronological consistency can be achieved without compromising performance. Our results show that ChronoBERT and ChronoGPT match or surpass the language comprehension abilities of BERT and GPT-2 while generating performance comparable to the much larger Llama 3.1 model in asset pricing applications.

Our findings reveal that the impact of lookahead bias in return prediction tasks is modest. We also highlight that the influence of lookahead bias is both model- and application-specific. Notably, downstream predictive models can adapt to limitations in language comprehension, ensuring economically and statistically significant gains. 

In addition to quantifying the impact of lookahead bias in return prediction using financial news, we propose a scalable framework for training chronologically consistent LLMs. This framework offers a constructive solution to deal with lookahead bias in predictive modeling, addressing a fundamental challenge in the application of LLMs to finance and other social sciences. By ensuring chronological consistency, our approach lays the foundation for more reliable applications of LLMs in these domains.

Our work suggests several potential avenues for future research. One avenue would be to assess lookahead bias in other settings. For this purpose, we make our ChronoBERT and ChronoGPT models publicly available at: \url{https://huggingface.co/manelalab}. Another avenue would be developing compute-optimal training strategies specifically tailored for chronologically consistent LLMs. While we have demonstrated that strong language understanding can be achieved without introducing lookahead bias, further work is needed to establish scaling laws (akin to Chinchilla scaling laws from \citet{hoffmann2022training}) that account for the unique constraints of temporal data limitations \citep{muennighoff2023scaling}. Such scaling laws would provide guidance on the optimal allocation of computational resources when training models with historical data cutoffs, potentially revealing different optimal ratios of parameters to training tokens compared to models trained on all available data. 

% This framework could help researchers determine the minimum model size needed to achieve competitive performance given a specific historical knowledge cutoff, or conversely, the amount of temporally filtered high-quality data needed for a given model architecture. These insights would be particularly valuable for applications in finance and economics where both historical accuracy and computational efficiency are paramount.

\clearpage
\newpage
\onehalfspacing
\bibliographystyle{jf}
\bibliography{lookahead_bias}

\newpage
\appendix
\counterwithin{figure}{section}
\renewcommand{\thefigure}{\thesection.\arabic{figure}}
\counterwithin{table}{section}
\renewcommand{\thetable}{\thesection.\arabic{table}}
\counterwithin{equation}{section}
\renewcommand{\theequation}{\thesection.\arabic{equation}}

\begin{center}{\bf{\LARGE Appendix}}\end{center}

\section{Language Understanding Evaluations}\label{sec:glue}

\subsection{GLUE Evaluation}

In this part, we lay out the details of the GLUE \citep{wang2018glue} evaluation process. Following \citet{warner2024smarter}, we use the same evaluation hyperparameters. Here are the details on learning rate, weight decay, and maximum number of epochs for each task. We use early stopping for all the fine-tuning tasks based on validation loss. The RTE, MRPC, and STS-B tasks are finetuned starting from the checkpoint of MNLI.

\begin{itemize}
    \item CoLA (Corpus of Linguistic Acceptability): learning rate: 8e-5; weight decay: 1e-6; maximum epochs: 5.
    \item SST-2 (Stanford Sentiment Treebank - Binary Classification): learning rate: 8e-5; weight decay: 1e-5; maximum epochs: 2.
    \item MNLI (Multi-Genre Natural Language Inference): learning rate: 5e-5; weight decay: 5e-6; maximum epochs: 1.
    \item MRPC (Microsoft Research Paraphrase Corpus): learning rate: 5e-5; weight decay: 5e-6; maximum epochs: 10.
    \item QNLI (Question Natural Language Inference): learning rate: 8e-5; weight decay: 5e-6; maximum epochs: 2. 
    \item QQP (Quora Question Pairs): learning rate: 5e-5; weight decay: 5e-6; maximum epochs: 10. 
    \item RTE (Recognizing Textual Entailment): learning rate: 5e-5; weight decay: 1e-5; maximum epochs: 3. 
    \item STS-B (Semantic Textual Similarity Benchmark): learning rate: 8e-5; weight decay: 5e-6; maximum epochs: 10.
\end{itemize}

\subsection{Hellaswag Evaluation}

To evaluate the ability of our ChronoGPT model to generate coherent and contextually appropriate text, we perform an autoregressive evaluation on the HellaSwag dataset. The goal is to assess how well the model assigns probabilities to different possible sentence completions, with the expectation that the most plausible completion receives the highest probability.

Given a sequence of input tokens \( x = (x_1, x_2, \dots, x_T) \), our model, parameterized by \( \theta \), produces a probability distribution over the vocabulary at each time step. The logit output for the sequence is given by:
$$
    \ell_t = f_{\theta}(x_1, \dots, x_{t-1}),
$$
where \( \ell_t \) represents the predicted logits for the token \( x_t \).

To compute the autoregressive loss, we shift the token sequence such that the model predicts each token based on previous tokens. The loss for each token is computed using the cross-entropy loss:
$$
    \mathcal{L}_{b,t} = -\log P_{\theta}(x_t \mid x_1, \dots, x_{t-1}),
$$
where:
$$
    P_{\theta}(x_t \mid x_1, \dots, x_{t-1}) = \frac{\exp(\ell_{t,x_t})}{\sum_{j} \exp(\ell_{t,j})}.
$$

Since the dataset consists of prompt + completion pairs, we ensure that the evaluation is performed only on the completion tokens. Given a binary mask \( M \) of shape \( (B, T) \), where \( M_{b,t} = 1 \) if token \( x_{b,t} \) belongs to the completion and \( 0 \) otherwise, the masked losses are:
$$
    \mathcal{L}_{\text{masked}, b, t} = M_{b,t} \cdot \mathcal{L}_{b,t}.
$$

To compute the total loss per sequence, we sum the loss of each individual completion token in the sequence:
$$
    L_{\text{sum}, b} = \sum_{t} \mathcal{L}_{\text{masked}, b, t}\,.
$$

The normalized loss per sequence can be calculated as:
$$
    L_{\text{avg}, b} = \frac{L_{\text{sum}, b}}{\sum_{t} M_{b,t}}\, ,
$$
where the denominator accounts for the number of valid completion tokens in each sequence.

Since each prompt is associated with multiple completions, we select the completion with the lowest normalized loss as the most probable one:
$$
    \hat{y} = \arg\min_{b} L_{\text{avg}, b}\,.
$$

This evaluation framework allows us to assess the model’s ability to rank plausible text completions by likelihood. A model with a lower average cross-entropy loss is expected to generate more coherent and contextually appropriate completions.

\newpage
\section{Scaling Law of ChronoGPT}\label{scaling_law_test}

A critical question is whether performance can be improved despite the inherent limitations of the chronological training data. According to scaling laws \citep{kaplan2020scaling}, model performance is a function of parameter count, data size, and compute budget. In our context, the latter two factors face a bottleneck. Scaling the training dataset is challenging due to the strict requirement of chronological consistency. Furthermore, with this constrained dataset, training beyond a certain point (e.g., four epochs) yields modest benefit on loss reduction.

Motivated by the success of model scaling demonstrated by OpenAI \citep{radford2019language} and AllenAI \citep{olmo20242}, we investigate if increasing model size is an effective strategy within our data-constrained regime. We test four architectures: 124M, 353M, 1B, and 1.5B parameters, detailed in Table \ref{tab:architecture}.

\begin{table}[!htb]
  \begin{center}
    \begin{tabularx}{\textwidth}{@{\hskip\tabcolsep\extracolsep\fill}lrrrrr}
      \toprule
      Model Size & Vocabulary & Layers & Heads & $d_{\text{model}}$ \\
      \midrule
      124 M  & 50,304 & 12 & 6  & 768  \\
      353 M  & 50,304 & 24 & 8  & 1,024 \\
      1 B    & 50,304 & 36 & 12 & 1,536 \\
      1.5 B  & 50,304 & 52 & 12 & 1,536 \\
      \bottomrule
    \end{tabularx}
  \end{center}
  \caption{Architecture hyperparameters for the four ChronoGPT model sizes.}
  \label{tab:architecture}
  \bigskip
  \small
  This table summarizes the core architectural hyperparameters of each scaled-up ChronoGPT model.
\end{table}

\begin{figure}[!htb]
    \begin{center}
    \includegraphics[width=\linewidth]{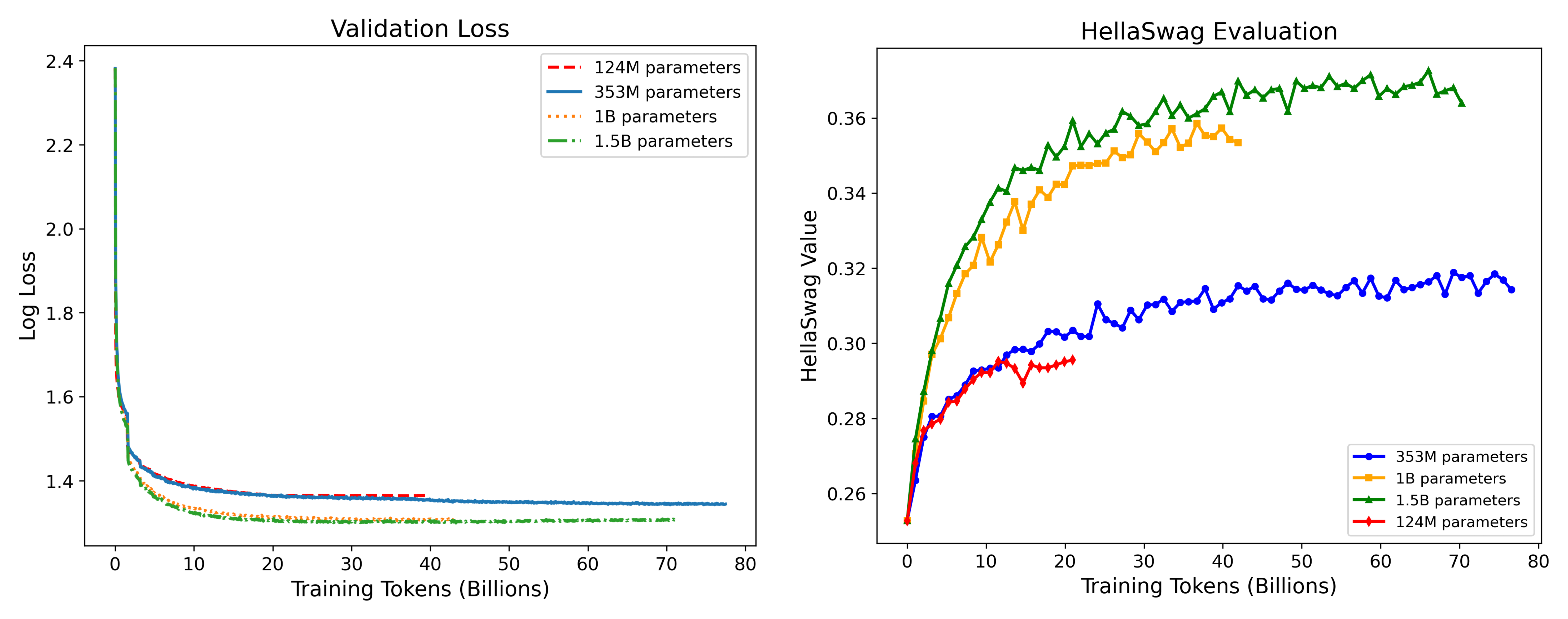}\\
    \end{center}
    \caption{Validation Loss and Language Understanding Performance Comparison}
    \label{fig:scaled_up}
    \bigskip
    \small
    This figure compares the cross-entropy loss and HellaSwag scores of ChronoGPT models with four sizes: 124 million (124M), 353 million (353M) parameters, 1 billion (1B) parameters, and 1.5 billion (1.5B) parameters. Validation losses are reported in logs. All models have a consistent knowledge cutoff of December 1999.
\end{figure}

We then evaluate them on validation loss and language understanding tasks against our baseline, using a consistent knowledge cutoff of December 1999. As shown in Figure \ref{fig:scaled_up}, larger models consistently achieve lower validation loss and higher HellaSwag scores, demonstrating that scaling up model size improves both general language modeling and specific language understanding capabilities. The 1.5B parameter model achieves the best performance across both metrics, with validation loss converging to approximately 1.32 after 80 billion training tokens, compared to 1.35 for the smallest 124M model. Similarly, the 1.5B model reaches a HellaSwag score of 0.37, substantially outperforming the 353M model (0.32) and the smallest 124M model (0.295).

These results confirm that scaling model size is an effective strategy for improving both language modeling and downstream task performance, even within a data-constrained regime. Based on these prominent gains and considering our compute budget, we select the 1.5B parameter architecture to train the full ChronoGPT series, extending its knowledge base from 1999 to 2024.

\clearpage
\newpage
\section{Architectural Differences Between BERT and GPT}\label{sec:architectures}

This section provides a concise overview of the fundamental architectural differences between BERT (Bidirectional Encoder Representations from Transformers) and GPT (Generative Pre-trained Transformer), explaining their distinct pretraining objectives and how they handle sequence classification tasks.

\subsection{Pretraining Objectives}

\subsubsection{BERT: Masked Language Modeling (MLM)}

BERT employs a bidirectional approach where it randomly masks some percentage of input tokens (30\% for ChronoBERT) and then predicts those masked tokens. For a given input sequence $X = (x_1, x_2, ..., x_n)$, some tokens are replaced with a special [MASK] token, creating a partially masked sequence $X_{masked}$. The pretraining objective is to predict the original tokens at the masked positions:
$$
\mathcal{L}_{MLM} = \mathbb{E}_{(X, m) \sim \mathcal{D}} \left[ -\sum_{i \in m} \log P(x_i | X_{masked}) \right],
$$
where $m$ is the set of masked token indices and $\mathcal{D}$ is the training distribution.

\subsubsection{GPT: Causal Language Modeling (CLM)}

GPT uses an autoregressive approach for pretraining, predicting the next token given all previous tokens in the sequence. For an input sequence $X = (x_1, x_2, ..., x_n)$, the model minimizes:
$$
\mathcal{L}_{CLM} = -\sum_{i=1}^{n} \log P(x_i | x_1, x_2, ..., x_{i-1}).
$$
This objective trains the model to generate coherent text by predicting each token based only on its preceding context.

\subsection{Attention Mechanisms}

The key architectural difference between BERT and GPT lies in their attention mechanisms:

\subsubsection{BERT: Bidirectional Self-Attention}

In BERT, each token attends to all tokens in the sequence, regardless of position. The self-attention computation for token $i$ at layer $\ell$ is:
$$
\text{Attention}(Q_i^{\ell}, K^{\ell}, V^{\ell}) = \text{softmax}\left(\frac{Q_i^{\ell} \cdot (K^{\ell})^T}{\sqrt{d_k}}\right) \cdot V^{\ell},
$$
where $Q_i^{\ell}$ is the query vector for token $i$, and $K^{\ell}$ and $V^{\ell}$ are the key and value matrices for all tokens in the sequence. This allows BERT to incorporate context from both before and after each token, capturing bidirectional relationships.

\subsubsection{GPT: Causal Self-Attention}

GPT uses a causal (or masked) self-attention mechanism where each token can only attend to itself and previous tokens:
$$
\text{CausalAttention}(Q_i^{\ell}, K^{\ell}, V^{\ell}) = \text{softmax}\left(\frac{Q_i^{\ell} \cdot (K^{\ell}_{1:i})^T}{\sqrt{d_k}}\right) \cdot V^{\ell}_{1:i}.
$$
This is achieved by masking future positions in the attention matrix, effectively limiting the receptive field to tokens $j \leq i$.

\subsection{Sequence Classification Approaches}

\subsubsection{BERT for Sequence Classification}

BERT performs sequence classification by utilizing the first token of the input sequence. After passing through the BERT transformer layers, the final hidden state of this token serves as a comprehensive representation of the entire sequence:
$$
h_{[CLS]} = \text{BERT}(x_1, x_2, ..., x_n)_{1}.
$$

A classification head (typically a linear layer followed by softmax) is then applied:
$$
P(y|X) = \text{softmax}(W \cdot h_{[CLS]} + b).
$$

During fine-tuning, the entire model, including the pretrained transformer and the classification head, is updated to minimize the cross-entropy loss:
$$
\mathcal{L}_{cls} = -\sum_{c=1}^{C} y_c \log(P(y_c|X)),
$$
where $C$ is the number of classes and $y_c$ is the ground truth label.

\subsubsection{GPT for Sequence Classification}

GPT handles sequence classification differently from BERT. Instead of using the first token, GPT leverages the last token as a representation of the entire sequence, as it attends to all preceding tokens. After passing through the GPT transformer layers, the final hidden state of the last token serves as a comprehensive sequence representation:
$$
h_{[CLS]} = \text{GPT}(x_1, x_2, ..., x_n)_{n}.
$$
Once the sequence representation is extracted, the classification head and loss functions remain the same as in BERT.

These architectural differences explain the performance variations observed between ChronoBERT and ChronoGPT in our experiments. While both models achieve chronological consistency, their underlying architectures make them suited to different types of tasks, with ChronoBERT excelling at classification and understanding benchmarks (GLUE) and ChronoGPT being better adapted to generative and completion tasks (HellaSwag).

\end{document}